\documentclass[%
 reprint,
 superscriptaddress,
 amsmath,amssymb,
 aps,
 prb,
]{revtex4-2}

\usepackage{physics}
\usepackage{graphicx}
\usepackage{dcolumn}
\usepackage{bm}
\usepackage{booktabs} 

\usepackage{xcolor}

\begin{document}

\title{Many-body perturbation theory vs. density functional theory:\\ A systematic benchmark for band gaps of solids}

\author{Max Großmann}
\email{max.grossmann@tu-ilmenau.de}
\affiliation{Institute of Physics and Institute of Micro- and Nanotechnologies, Technische Universit\"at Ilmenau, 98693 Ilmenau, Germany}
\affiliation{These authors contributed equally to this work.}

\author{Marc Thieme}
\affiliation{Institute of Physics and Institute of Micro- and Nanotechnologies, Technische Universit\"at Ilmenau, 98693 Ilmenau, Germany}
\affiliation{These authors contributed equally to this work.}

\author{Malte Grunert}
\affiliation{Institute of Physics and Institute of Micro- and Nanotechnologies, Technische Universit\"at Ilmenau, 98693 Ilmenau, Germany}

\author{Erich Runge}
\affiliation{Institute of Physics and Institute of Micro- and Nanotechnologies, Technische Universit\"at Ilmenau, 98693 Ilmenau, Germany}

\date{\today}

\begin{abstract}
We benchmark many-body perturbation theory against density functional theory (DFT) for the band gaps of solids.
We systematically compare four $GW$ variants---$G_{0}W_{0}$ using the Godby-Needs plasmon-pole approximation ($G_{0}W_{0}$-PPA), full-frequency quasiparticle $G_{0}W_{0}$ (QP$G_{0}W_{0}$), full-frequency quasiparticle self-consistent $GW$ (QS$GW$), and QS$GW$ augmented with vertex corrections in $W$ (QS$G\hat{W}$)---against the currently best performing and popular density functionals mBJ and HSE06.
Our results show that $G_{0}W_{0}$-PPA calculations offer only a marginal accuracy gain over the best DFT methods, however at a higher cost.
Replacing the PPA with a full-frequency integration of the dielectric screening improves the predictions dramatically, almost matching the accuracy of the QS$G\hat{W}$.
The QS$GW$ removes starting-point bias, but systematically overestimates experimental gaps by about $15\%$.
Adding vertex corrections to the screened Coulomb interaction, i.e., performing a QS$G\hat{W}$ calculation, eliminates the overestimation, producing band gaps that are so accurate that they even reliably flag questionable experimental measurements.
\end{abstract}

\maketitle

\section*{Introduction}

Ever since the dawn of condensed matter physics, the band gap of semiconductors and insulators has been one of the most important material properties for optical and optoelectronic applications \cite{Shockley1950}. 
Despite its relative ease of measurement, accurately predicting the band gap from first principles remains a challenging task, especially because interpreting the Kohn-Sham (KS) gap from density functional theory (DFT), the workhorse of theoretical materials science, as the fundamental band gap leads to a systematic underestimation of the band gaps of solids \cite{Perdew1985}.

A benchmark by Borlido~\textit{et al.}~\cite{Borlido2019, Borlido2020} consisting of 472 non-magnetic materials, evaluating the performance of 21 DFT functionals, showed that functionals from the upper rungs of Perdew's "Jacob's ladder" \cite{Perdew2001}, such as meta-generalized gradient approximation (meta-GGA) functionals and hybrid functionals, can significantly reduce the systematic underestimation of the band gap.
However, these improvements are often due to \mbox{(semi-)empirical} adjustments and not the result of a solid theoretical basis \cite{Borlido2020}.

In contrast to DFT, many-body perturbation theory (MBPT) offers a fundamentally different way to obtain accurate electronic structures of solids. 
Based on a rigorous diagrammatic expansion of the electron correlation, MBPT provides, in principle, a systematic way to improve accuracy by incorporating higher-order corrections \cite{Hedin1965, Onida2002}.
Due to methodological advances over the last decades, in particular different "flavors" and implementations of the so-called $GW$ approximation to the electronic self energy \cite{Shishkin2006, Shishkin2007, Kotani2007, Shishkin2007_vertex, Liu2016, Kutepov2020, Leon2021, Kutepov2022, Leon2023, Cunningham2023}, and in part due to ever-increasing computational power, MBPT has evolved from a niche method to an attractive practical tool for accurate calculations of band gaps of solids.
However, we must note that MBPT, as the name implies, is a perturbative method typically starting from a DFT-derived electronic structure, and is therefore generally more expensive to perform than a DFT calculation.

Despite its strong theoretical foundation and the common belief that MBPT has superior accuracy compared to state-of-the-art DFT functionals, a systematic large-scale benchmark directly comparing MBPT with state-of-the-art DFT functionals for the band gap of solids remains to be performed.

The demand for a systematic benchmark is particularly pressing in light of recent trends in materials science: the rise of machine learning (ML) and the need for accurate datasets \cite{Butler2018}. 
ML models, which have rapidly become indispensable for accelerating the discovery of new materials and the fast prediction of key properties ranging from the transition temperature of conventional superconductors \cite{Stanev2018, Cerqueira2023, Sanna2024} and diffusion coefficients \cite{Grunert2025a, Angeletti2025} to optical spectra \cite{Ibrahim2024, Hung2024, Grunert2024b}, depend on training data, typically generated by DFT.
Given the limitations of these DFT datasets, there is a steadily growing concern about the reliability of ML predictions, especially as models become more sophisticated and accurate.
A promising solution to these concerns is transfer learning \cite{Hoffmann2023, Grunert2025b}, which requires a (small) dataset of high-fidelity data for model retraining.
Therefore, a benchmark of MBPT against DFT is also invaluable for the ML part of the materials science community, as one needs to decide on the method of choice when computing expensive datasets for transfer learning.

Against this backdrop, our work provides a systematic benchmark of MBPT against the best meta-GGA and hybrid DFT functionals for calculating the band gap of solids, and answers the central question: \textit{"How well do MBPT calculations perform relative to the best available DFT functionals when predicting the band gaps of solids?"}.

The basis for our MBPT benchmark is the earlier DFT benchmark of Borlido~\textit{et al.}~\cite{Borlido2019, Borlido2020}.
We have adopted their extensive dataset of experimental band gaps for 472 non-magnetic semiconductors and insulators, using experimental crystal structures and geometries from the ICSD \cite{Bergerhoff1983, Belsky2002, Zagorac2019} to facilitate direct a comparison.
For details on the curation of the experimental data, as well as a detailed analysis of the dataset in terms of contained elements and band gap distribution, we refer to the original publications \cite{Borlido2019, Borlido2020}.
The dataset of Borlido~\textit{et al.}~\cite{Borlido2019, Borlido2020} includes not only experimental band gaps, but also band gaps calculated using a wide range of 21 DFT functionals, from which we have selected the best performing meta-GGA functional (mBJ) \cite{Tran2009}, and the best performing hybrid functional (HSE06) \cite{Heyd2003, Heyd2006} for comparison.
The selected functionals not only present the upper rungs of Perdew's "Jacob's ladder", ensuring a broad and representative coverage, but also include one of the most widely used functionals in condensed matter physics, HSE06.

As the MBPT method of choice, we used the $GW$ approximation, the arguably most successful and widely used MPBT method \cite{Reining2018}.
Because the $GW$ approximation exists in numerous variants, implementations, and workflows, it is impractical to cover every published variant exhaustively. 
Consequently, we have limited ourselves to a strategically chosen subset of four methods:
(i) One-shot $GW$, so called $G_{0}W_{0}$ calculations, utilizing the Godby-Needs plasmon-pole approximation (PPA) \cite{Godby1989}.
(ii) Full-frequency quasiparticle $G_{0}W_{0}$ (QP$G_{0}W_{0}$) calculations.
(iii) Full-frequency quasiparticle self-consistent $GW$ (QS$GW$) calculations.
(iv) QS$GW$ calculations with vertex corrections in the screened Coulomb interaction $W$ (QS$G\hat{W}$).

All calculations were automated using custom, in-house workflows to ensure reproducible results, see Methods section.
The workflow codes and generated data can be found in the Data and Code Availability section.
The starting point for all calculations was an LDA \cite{Marques2012} DFT calculation.
For method (i), we also repeated all $G_{0}W_{0}$ calculations starting from a PBE \cite{Marques2012} DFT calculation to investigate the influence of the starting point.
The DFT calculations in method (i) are based on norm-conserving pseudopotentials using a plane-wave basis, while those in methods (ii--iv) are all-electron calculations using a linear muffin-tin orbital (LMTO) basis.
Calculations for (i) were performed using \textsc{Quantum ESPRESSO} \cite{Giannozzi2009all, Giannozzi2017all} and \textsc{Yambo} \cite{Marini2009, Sangalli2019AllAuthors}.
Those for (ii--iv) were performed using the \textsc{Questaal} code \cite{Kotani2007, Pashov2020, Cunningham2023}.
Details regarding the workflows, convergence parameters, and other computational choices are reported in the Methods section.

The four $GW$ schemes were selected to represent a hierarchy of computational cost, physical rigor, and methodological maturity in MBPT.
Method (i) is a widely used and comparatively inexpensive variant of the $GW$ approximation, implemented in most plane-wave pseudopotential MBPT codes \cite{Hybertsen1986, Deslippe2012, Marini2009, Sangalli2019AllAuthors, Mortensen2024}. 
Here, the quasiparticle (QP) energies $\epsilon_i^\text{QP}$ are calculated from the KS energies $\epsilon_i^\text{KS}$ via a linearized QP equation:
\begin{equation} \label{eq:linqp}
    \epsilon_{i}^\text{QP} = \epsilon_{i}^\text{KS} + Z_{i}\,\big\langle \phi_i^\text{KS}\big\vert \,\big(\Sigma(\epsilon_{i}^\text{KS}) - V_{xc}^\text{KS}\big)\,\big\vert \phi_i^\text{KS}\big\rangle
\end{equation}
neglecting the off-diagonal elements of the self energy $\Sigma$.
Here, $Z_{i}^{-1} = 1\,-\,\langle \phi_i^\text{KS}\vert \,\,d\Sigma/\mathrm{d}\epsilon|_{\epsilon_{i}^\text{KS}}\,\,\vert \phi_i^\text{KS}\rangle$ is the renormalization factor, $V_{xc}^\text{KS}$ is the KS exchange-correlation potential, and $\vert \phi_i^\text{KS}\rangle$ are KS states \cite{Onida2002}.
Methods (ii--iv) "quasiparticlize" the energy-dependent $\Sigma$ by constructing a
static Hermitian potential from it \cite{Schilfgaarde2006}
\begin{equation}
    \Sigma_0 = \frac{1}{2}\sum_{ij} \vert\psi_i\rangle
    \{
    \Re[\Sigma(\epsilon_i\bigr)]_{ij} + \Re[\Sigma(\epsilon_j)]_{ij}
    \}
    \langle\psi_j\vert
    \label{eq:qp}
\end{equation}
replacing $V_{xc}^\text{KS}$.
QP energies and wavefunctions are then obtained by solving the resulting effective KS equations, in which $V_{xc}^\text{KS}$ is replaced by $\Sigma_0$.
Thereby, the off-diagonal elements of $\Sigma$ are included, which turn out to be necessary to obtain correct band topologies \cite{Schilfgaarde2006}, since the LDA incorrectly orders bands for, e.g., simple narrow-gap semiconductors like InN \cite{Cunningham2023} and PbTe \cite{Pashov2020}.
The $GW$ variant (ii) uses an all-electron starting point and replaces the PPA of the frequency dependence of the dielectric screening with a numerical integration that captures its exact frequency dependence. 
This is referred to as a full-frequency $GW$ calculation and is also used in methods (iii) and (iv).
The self-consistent QP$GW$ scheme (iii), the QS$GW$, further eliminates the dependence on the initial DFT reference by iterating the quasiparticle Eq.~(\ref{eq:qp}) to convergence, i.e., until $\Sigma_0^\text{in} = \Sigma_0^\text{out}$ \cite{Pashov2020}.
In detail, one performs the following loop \cite{Pashov2020}:
\begin{equation}
   \dots \rightarrow \Sigma_0 \rightarrow G_0 \rightarrow \Sigma = iG_0 W[G_0] \rightarrow \Sigma_0 \rightarrow \dots
\end{equation}
Here, $\Sigma_0$ is again obtained from Eq.~(\ref{eq:qp}) and $G_0$ is obtained from the solution of the effective KS equations, in which $V_{xc}^\text{KS}$ is replaced by $\Sigma_0$.
Finally, method (iv), the QS$G\hat{W}$ introduced in the seminal work by Cunningham \textit{et al.}~\cite{Cunningham2023}, was chosen because it is, to our knowledge, the most advanced practical $GW$ method.
It has been found to produce extremely accurate band gaps for 43 more or less well understood semiconductors, as well as for some strongly correlated antiferromagnetic oxides.
The QS$G\hat{W}$ method accomplishes this by incorporating vertex corrections into the screened Coulomb interaction $W$.
In particular, vertex corrections that augment dielectric screening are obtained by solving the Bethe-Salpeter equation (BSE) explicitly, thereby incorporating excitonic effects.
In the employed \textsc{Questaal} implementation \cite{Cunningham2023}, the BSE is solved for every momentum-transfer vector $\mathbf{q}$ in the k-point grid explicitly, using the usual static approximation of the BSE kernel in the Tamm-Dancoff approximation (TDA).
Consequently, each QS$G\hat{W}$ iteration requires the computation of the full excitonic dispersion, resulting in a significant increase in computational cost compared to a QS$GW$ calculation.
At this point, we note that even more advanced implementations of vertex corrections and $GW$ schemes already exist, as also stated by Cunningham~\textit{et al.}~\cite{Cunningham2023}.
For example, Refs.~\cite{Bruneval2005, Shishkin2007, Franchini2010} included ladder diagrams through an effective, nonlocal, static kernel constructed within time-dependent DFT to mimic the BSE.
Kutepov~\cite{Kutepov2016, Kutepov2017, Kutepov2022} even proposed several schemes for self-consistently solving Hedin’s equations, including vertex corrections, which further improve the results for some materials exhibiting complex electronic correlation effects.
However, from a practical point of view, current computational limitations restrict us from applying them to a benchmark dataset of this scale.
We expect that this will change in the coming years.

The work of Cunningham \textit{et al.}~\cite{Cunningham2023} was a first preliminary benchmark of the QS$G\hat{W}$ method. 
Here, we present a systematic extension of it for a wider variety of materials.
In doing so, we also validate previous results, a practice that unfortunately is still uncommon in modern computational materials science.

Throughout this study, we treat the QS$G\hat{W}$ band gaps as our reference standard, using them to benchmark other \textit{ab initio} methods and to identify questionable experimental values.

Despite our original ambition to benchmark all 472 materials contained in the dataset, the steep computational scaling of $GW$ calculations---formally $\mathcal{O}(N^4)$ for conventional implementations, where $N$ is the atoms in the unit cell---makes materials with larger unit cells prohibitively expensive.
Since the benchmark dataset contains materials up to $N=28$, we are forced to introduce a system size limit.
For method (i), we have performed calculations for materials with a maximum of 12 atoms per unit cell and atomic numbers up to 83 (excluding the lanthanides due to unavailable pseudopotentials).
We have successfully converged and obtained band gaps for 286 (301) out of 332 systems using method (i) starting from an LDA (PBE) DFT calculation.
The more demanding all-electron $GW$ variants \{(ii), (iii), (iv)\} were further restricted to unit cells of six atoms or less, but no additional element restriction was imposed.
There, we obtained converged band gaps for \{156, 154, 101\} of 171 materials.

The remaining compounds that failed to converge did so for technical reasons only.
This mostly happened when we had insufficient memory to run the calculation or when the available CPU time was insufficient.
High memory requirements often occurred for materials that required dense k-point grids to converge the band gap.
In some cases, memory requirements were also high due to low unit cell symmetry, which indirectly increased the number of k-points used in a calculation.
In other cases, it was due to the large number of electrons inherent in all-electron calculations and pseudopotentials containing many semicore states, e.g., Ag and Hg. 
In method (iv), the number of matrix elements of the Bethe-Salpeter Hamiltonian to be stored in memory scales as $(N_v \times N_c \times N_k)^2$, where $N_v$ and $N_c$ are the number of valence and conduction included in the BSE transition space, respectively, and $N_k$ is the number of k-points. 
Therefore, if materials require either a dense k-point grid, such as Ge, or many bands, such as HgI, the memory requirements become enormous when vertex corrections in $W$ are included in the self-consistency loop.
This further limits the number of materials that we were able to calculate using method (iv).
Despite these convergence and memory bottlenecks, we still ran many thousands of $GW$ computations to obtain converged results, using several million CPU hours in total and up to several hundred gigabytes of memory per calculation, highlighting the steep scaling of these MBPT methods.

We have provided a spreadsheet containing the band gaps for all calculated materials as additional Supplementary Material, available online.
For the complete dataset, including band structures and densities of states for all materials calculated using methods (ii--iv), refer to the Data and Code Availability section.

\section*{Results}

\begin{table*}[ht]
    \caption{
    \textbf{Error metrics for all investigated calculation methods.}
    The following metrics are given for the calculated band gaps using the methods described in the text: Dataset size ($n$), mean error (ME), standard deviation of the error ($\sigma$), mean absolute error (MAE), root mean square error (RMSE), and mean absolute percentage error (MAPE).
    The data for mBJ and HSE06 are taken from Borlido \textit{et al.}~\cite{Borlido2019, Borlido2020}.
    The error is defined as $E_{\mathrm{gap}}^{\mathrm{calc.}} - E_{\mathrm{gap}}^{\mathrm{exp.}}$.
    The metrics in the upper block are based on all available calculations for each method, as indicated by the $n$ row. 
    The lower block, on the other hand, restricts the comparison to the 94 materials for which data is available for all methods.
    Note that the MAPE has a tendency to preference methods which systematically underestimate band gaps, see explanation in the text.
    The best method for each metric is highlighted in bold.\\
    }
    \centering
    \begin{tabular}{lccccccccc}
    \toprule
     & mBJ & HSE06 & $G_{0}W_{0}$@LDA-PPA & $G_{0}W_{0}$@PBE-PPA & QP$G_{0}W_{0}$ & QP$G_{0}W_{0}$+SOC & QS$GW$ & QS$G\hat{W}$ & QS$G\hat{W}$+SOC \\
    \midrule    
    $n$ & 471 & 472 & 286 & 301 & 156 & 153 & 154 & 101 & 96 \\[1mm]
    ME (eV) & -0.22 & -0.10 & -0.03 & -0.12 & 0.08 & \textbf{-0.01 }& 0.55 & 0.14 & 0.05 \\[1mm]
    $\sigma$ (eV) & 0.68 & 0.85 & 0.75 & 0.82 & 0.54 & 0.55 & 0.63 & \textbf{0.45} & 0.47 \\[1mm]
    MAE (eV) & 0.50 & 0.53 & 0.54 & 0.60 & 0.37 & 0.37 & 0.64 & 0.31 & \textbf{0.30} \\[1mm]
    RMSE (eV) & 0.72 & 0.86 & 0.75 & 0.83 & 0.55 & 0.55 & 0.83 & \textbf{0.47} & \textbf{0.47} \\[1mm]
    MAPE (\%) & 30 & 31 & 39 & 40 & 32 & 26 & 48 & 29 & \textbf{21} \\[1mm]
    \midrule
    \\[-3mm]
    $n$ & \multicolumn{9}{c}{Materials for which experimental and computational data are available for all methods: 94} \\[1mm]
    ME (eV) & -0.42 & -0.68 & -0.13 & -0.23 & 0.06 & \textbf{-0.03} & 0.56 & 0.13 & 0.04 \\[1mm]
    $\sigma$ (eV) & 0.68 & 1.16 & 0.57 & 0.68 & 0.46 & 0.47 & 0.54 & \textbf{0.45} & 0.47 \\[1mm]
    MAE (eV) & 0.58 & 0.82 & 0.46 & 0.57 & 0.31 & 0.31 & 0.60 & 0.30 & \textbf{0.29} \\[1mm]
    RMSE (eV) & 0.80 & 1.34 & 0.58 & 0.71 & 0.47 & 0.47 & 0.78 & 0.47 & \textbf{0.46} \\[1mm]
    MAPE (\%) & 35 & 38 & 35 & 38 & 29 & \textbf{20} & 42 & 30 & 21 \\[1mm]
    \bottomrule
    \end{tabular}
    \label{tab:metrics}
\end{table*}

\begin{figure*}[ht]
    \centering
    \includegraphics{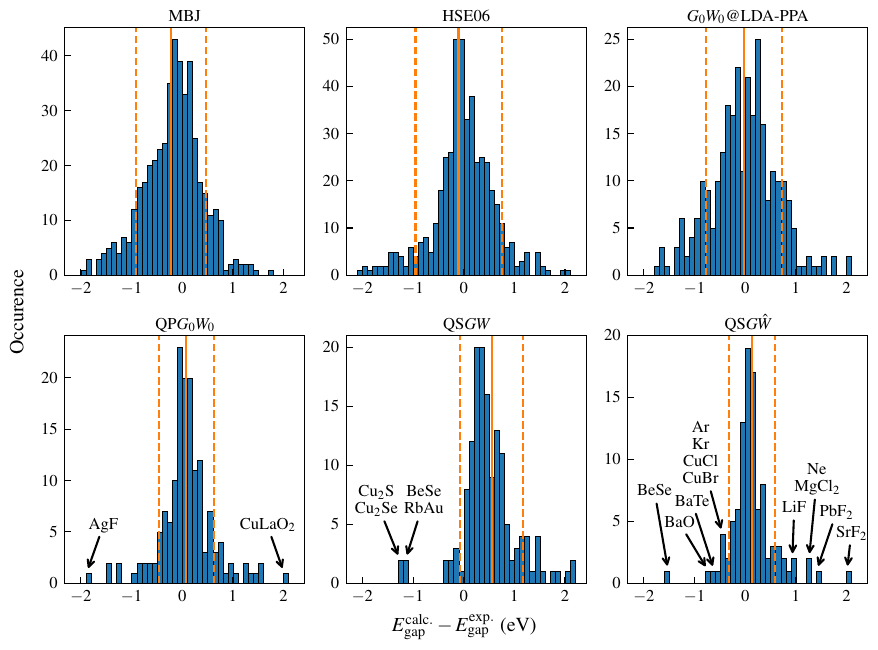}
    \caption{
    \textbf{Histograms showing the differences between calculated and experimental band gaps.}
    Each bin has a width of $0.1$~eV.
    The mean error (ME) is indicated by a solid orange line, while the standard deviation ($\sigma$) is indicated by dashed orange lines.
    Arrows in the histograms for the QP$G_{0}W_{0}$, QS$GW$, and QS$G\hat{W}$ calculations mark outliers which will be analyzed in detail in the text.
    }
    \label{fig:histo}
\end{figure*}

In the following, the four $GW$ methods (i--iv) introduced above will be referred to as $G_{0}W_{0}$-PPA, QP$G_{0}W_{0}$, QS$GW$, and QS$G\hat{W}$, respectively.
All $G_{0}W_{0}$-PPA gaps are converged to within $25$~meV or better, while the more
demanding full-frequency and self-consistent variants (ii--iv) are converged to within $100$~meV or better, see Methods for details.

Before discussing the benchmark results, we first want to mention the fact that, depending on the DFT implementation used, the starting point and band gap of our $GW$ calculation differed slightly from those reported in the original benchmark dataset \cite{Borlido2019, Borlido2020}. 
In some cases, materials classified as narrow-gap semiconductors in LDA calculations with a plane-wave basis set and projector augmented wave (PAW) pseudopotentials in the original benchmark dataset are metals in our norm-conserving pseudopotential and all-electron calculations.
We analyze this in detail in Supplementary~Note~1 of the Supplementary Information and attribute this to the fact that the PAW pseudopotential can be "softer" than the norm-conserving pseudopotential (and all-electron potentials).
Deviations are more pronounced when two pseudopotentials, created using different methodologies, include different semicore states for elements such as Cu, Ag, Se, and Hg.

\begin{figure*}[ht]
    \centering
    \includegraphics{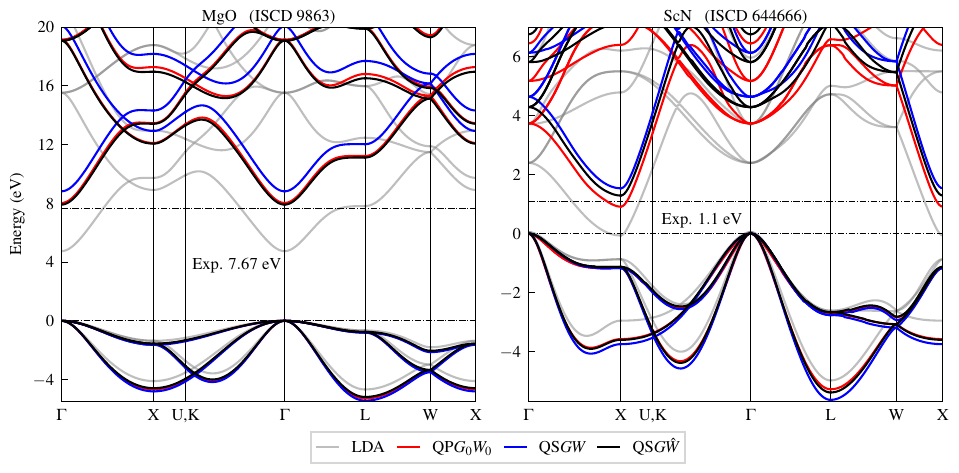}
    \caption{
    \textbf{Band structures calculated using different methods for the exemplary materials MgO (left panel) and ScN (right panel).}
    The dashed horizontal lines mark the experimental band gaps.
    The band gap for MgO is taken from Ref.~\cite{Madelung2004}, and the band gap for ScN is taken from Grümbel \textit{et al.}~\cite{Gruemble2024}, as discussed at the end of the Results section.
    The structures used for the calculations are identified by their respective ICSD database identifiers \cite{Bergerhoff1983, Belsky2002, Zagorac2019}, which are given in the panel titles.
    }
    \label{fig:bs}
\end{figure*}

\subsection*{Benchmark of computational methods}

To start, we benchmark all selected computational methods against experimental data using several statistical error metrics, which are summarized in Tab.~\ref{tab:metrics}.
As is customary in benchmarks, we report the mean error (ME), the standard deviation of the error ($\sigma$), the mean absolute error (MAE), the root mean squared error (RMSE), and the mean absolute percentage error (MAPE).
Here, the error is defined as $E_{\mathrm{gap}}^{\mathrm{calc.}} - E_{\mathrm{gap}}^{\mathrm{exp.}}$.
We also investigated the effect of spin-orbit coupling (SOC) on the QP$G_{0}W_{0}$ and QS$G\hat{W}$ results (see Methods for details), referred to as QP$G_{0}W_{0}$+SOC and QS$G\hat{W}$+SOC, respectively, in Tab.~\ref{tab:metrics}.
Histograms depicting the difference between computed and experimental band gaps for selected methods are presented in Fig.~\ref{fig:histo}.
For the sake of completeness, the histograms for the remaining methods are provided in Supplementary~Note~2.
Materials for which a method underestimates (overestimates) the experimental band gap are shown on the negative (positive) side of the histograms.
We marked the ME and $\sigma$ with solid and dashed orange lines, respectively.

Before analyzing the benchmark results, please note that the derived error metrics and histograms are for all materials for which a corresponding calculation is available.
For example, Borlido \textit{et al.}~\cite{Borlido2019, Borlido2020} obtained band gaps for all 472 materials, whereas we "only" performed QS$G\hat{W}$ calculations for 101 materials, as stated in Tab.~\ref{tab:metrics}.
However, we also calculated all error metrics for a subset of 94 materials for which data is available for all investigated methods to facilitate a more direct comparison (cf.~lower block of Tab.~\ref{tab:metrics}).

Regardless of the dataset used to evaluate the error metrics, the QS$G\hat{W}$ consistently emerges as the most accurate method, followed closely by the QP$G_{0}W_{0}$. 
Both mBJ and HSE06 tend to underestimate the experimental band gaps and exhibit high MAE and RMSE.
$G_{0}W_{0}$@LDA-PPA calculations outperform $G_{0}W_{0}$@PBE-PPA calculations.
Perhaps somewhat unexpectedly, the $G_{0}W_{0}$@LDA-PPA and $G_{0}W_{0}$@PBE-PPA calculations perform worse than mBJ in virtually all metrics when evaluated on the full datasets, except for the ME, where mBJ shows a systematic underestimation of the experimental band gap.
While $G_{0}W_{0}$@LDA-PPA calculations perform better than HSE06 when evaluated on the full datasets, $G_{0}W_{0}$@PBE-PPA calculations do not.
However, when the metrics of mBJ, HSE06, $G_{0}W_{0}$@LDA-PPA, and $G_{0}W_{0}$@PBE-PPA are compared on the subset of 94 materials for which data is available for all the investigated methods, the opposite is observed.
Here, both $G_{0}W_{0}$-PPA methods outperform mBJ and HSE06, as the metrics for mBJ and HSE06 worsen dramatically, while those for the $G_{0}W_{0}$-PPA methods improve.
No matter which dataset is chosen, the performance of mBJ and HSE06 is still impressive, considering that they are much less computationally demanding and easier to converge than $GW$ calculations \cite{Grossmann2024}.
The QS$GW$ lags far behind the QP$G_{0}W_{0}$ and the QS$G\hat{W}$ due to its well-known tendency to systematically overestimate band gaps by about $15\%$ (cf.~Ref~\cite{Cunningham2023}), and performs the worst overall in terms of error metrics.
The inclusion of spin-orbit coupling (SOC) in the QP$G_{0}W_{0}$ and QS$G\hat{W}$ calculations reduces the ME, bringing the calculated band gaps closer to the experimental values, modestly improves the ME and MAPE, but leaves other error metrics virtually unchanged.

At this point, we need to discuss and highlight an issue with the MAPE, a metric which is commonly used for band gap benchmarking.
Firstly, the MAPE for mBJ and HSE06 is much better than the MAPE of the $G_{0}W_{0}$-PPA methods and close to the MAPE of the QS$G\hat{W}$.
Secondly, the MAPE improves significantly when SOC is included in the QP$G_{0}W_{0}$ and the QS$G\hat{W}$ calculations.
However, this is apparently contradictory to the other metrics and highlights a known limitation of MAPE: it systematically favors methods that underestimate band gaps \cite{Tofallis2015}. 
Furthermore, we note that the MAPE is highly affected by materials with small band gaps, where relatively small changes in the calculated band gap lead to extremely large percentage errors, see Supplementary~Note~3.
Therefore, the distribution of band gaps in a dataset (i.e., the presence of more or fewer materials with smaller or larger band gaps) affects this metric heavily.
Consequently, we argue that MAPE should not be relied upon when evaluating the accuracy of computational methods for band gap prediction.
We suggest using the MAE and the RMSE as primary metrics in the future, as is commonly done in the machine learning part of the materials science community.

We will now elaborate further on these findings and explain them in more detail using the histograms shown in Fig.~\ref{fig:histo}, following the hierarchy of methodological complexity introduced earlier. 

We begin by explaining why changing the starting point of a $G_{0}W_{0}$-PPA from LDA to PBE worsens all error metrics (see Tab.~\ref{tab:metrics}).
At first, we were somewhat surprised by this, as our observations and experience suggest that $G_{0}W_{0}$@PBE-PPA calculations are used the more frequently in practice than $G_{0}W_{0}$@LDA-PPA calculations.
However, the superior performance of the LDA starting point compared to PBE for $G_{0}W_{0}$ calculations can be explained by the fact that $G_{0}W_{0}$ calculations rely heavily on fortuitous error cancellations, as discussed in detail in Ref.~\cite{Cunningham2023}.
This is usually rationalized as follows:
The band gap and the dielectric screening are inversely correlated \cite{Carrico2024}. 
LDA underestimates band gaps by about $50\%$ \cite{Borlido2019, Borlido2020} and thus overestimates the dielectric screening, while the random phase approximation (RPA)  \cite{Bohm1951}, used to calculate the screened Coulomb interaction $W$, systematically underestimates dielectric screening in semiconductors and insulators due to the absence of excitonic effects \cite{Cunningham2023}. 
In $G_{0}W_{0}$@LDA-PPA calculations, the LDA-induced over-screening and the RPA-induced under-screening compensate in a fortuitous way, leading to band gaps in good agreement with experiment. 
Changing the starting point from LDA to PBE reduces this compensation because PBE predicts larger gaps \cite{Borlido2019, Borlido2020} and hence weaker screening than the LDA, leading to slightly worse $G_{0}W_{0}$ band gaps.
This finding implies that the optimal $G_{0}W_{0}$ starting point is not simply "the most modern GGA" but the one whose self-consistent eigenvalues and screening happen to balance the systematic errors of the one-shot correction.
That said, it remains plausible that beginning with a DFT functional from the upper rungs of Perdew's "Jacob's ladder" \cite{Perdew2001} (such as PBE0 or HSE06) could deliver better one-shot quasiparticle gaps, as hinted at in Ref.~\cite{Gant2022}.
However, there is a clear trade-off: each rung above PBE increases the cost of the ground-state calculation considerably, and some functionals can be nearly as costly as the $G_{0}W_{0}$ step itself.

Compared to the $G_{0}W_{0}$-PPA calculations, the QP$G_{0}W_{0}$ is in much better agreement with the experiment and performs almost as well as the QS$G\hat{W}$.
Since three parts of the QP$G_{0}W_{0}$ calculation method have changed compared to the $G_{0}W_{0}$-PPA, we disentangle their individual impacts in the following:
Firstly, we suspect that most of the improvements are due to the replacement of the PPA with a full-frequency integration of the dielectric screening.
Secondly, since the starting point band gaps obtained by norm-conserving pseudopotential and all-electron calculations agree well, cf.~Supplementary~Note~1, the effect of the all-electron DFT starting point appears minor for most of the considered materials.
Thirdly, the substitution of the linearized quasiparticle Eq.~(\ref{eq:linqp}) for the effective potential given by Eq.~(\ref{eq:qp}) and following solution of the resulting KS equation most strongly affects materials where the LDA incorrectly ordered bands. 
Therefore, on average, this should also have a minor impact for most of the considered materials.

For the QS$GW$, we confirm the well-known fact that self-consistent RPA-based $GW$ schemes that omit vertex corrections systematically overestimate band gaps---by about $15\%$ in our benchmark.
This is reflected in the much larger ME of $0.55$~eV compared to the ME of $0.08$~eV for the QP$G_{0}W_{0}$.
The origin of this systematic overestimation can be traced back to the error cancellation described above. 
The inclusion of self-consistency removes the dependence of the results on the starting point, thus correcting the initial band gap error of the DFT starting point.
When $W$ is then evaluated in the RPA, the dielectric screening is systematically underestimated due to missing excitonic effects, leading to band gaps that are systematically too large.

The QS$G\hat{W}$ alleviates this underestimation of the dielectric screening by including vertex corrections in $W$, i.e., excitonic effects.
Looking at the corresponding histogram in Fig.~\ref{fig:histo}, we see that the QS$G\hat{W}$ does not improve the ME compared to the QP$G_{0}W_{0}$, but it reduces $\sigma$ considerably from $0.54$~eV to $0.45$~eV.
However, when comparing the QS$G\hat{W}$ and QP$G_{0}W_{0}$ on the subset of 94 materials for which data is available for all the investigated methods, we observe that both perform about equally well with respect to all error metrics except the ME.

The band structures of the exemplary materials MgO and ScN, shown in Fig.~\ref{fig:bs}, illustrate the points made above.
For MgO (left panel), the QS$G\hat{W}$ (black lines) and QP$G_{0}W_{0}$ (red lines) are nearly identical.
For both methods, the band gap agrees well with the experimental value.
The systematic band gap overestimation of the QS$GW$ is evident in all conduction bands.
For ScN (right panel), however, the QP$G_{0}W_{0}$ (red lines) deviates from the QS$G\hat{W}$ (black lines), by about the same amount as the QS$GW$, but in the opposite direction.
The QS$G\hat{W}$ band gap agrees well with experimental data.
This demonstrates that, while the QP$G_{0}W_{0}$  relies on a fortuitous cancellation of errors that can produce electronic structures in line with the much more expensive QS$G\hat{W}$ on average (cf.~Tab.~\ref{tab:metrics}), its accuracy for specific materials may vary.
Note also that the LDA band structure (gray lines) for ScN is that of a (semi-)metal.

The inclusion of SOC in the QP$G_{0}W_{0}$ and QS$G\hat{W}$ significantly improves the calculated band gap of materials containing heavy elements from the lower parts of the periodic table.
This is consistent with the fact that the spin-orbit coupling constant scales approximately with $Z^4$ in the atomic limit.
A prime example is PbTe, where the inclusion of SOC reduces the QS$G\hat{W}$ band gap from $0.89$~eV to $0.34$~eV, bringing it much closer to the experimental value of $0.19$~eV \cite{Madelung2004}. 
Similar improvements are observed for other heavy element compounds such as SnTe, CdTe, and PbSe, cf.~Supplementary~Material. 
These results highlight the importance of including SOC when aiming for high accuracy in heavy-element materials.

\subsection*{Investigating outliers}

Since QP$G_{0}W_{0}$ and QS$G\hat{W}$ calculations proved to be highly accurate, we examined materials with significant deviations from the reported experimental band gaps in detail.
Additionally, we investigated materials for which the QS$GW$ significantly underestimates the experimental band gap.
In other words, we examined cases that contradict the tendency of the QS$GW$ to overestimate band gaps.
Identified outliers are highlighted in Fig.~\ref{fig:histo} with arrows in the associated histograms.
These are often "unusual" compounds, such as an alkali auride.
Some outliers are caused by the absence of electron-phonon interactions in the calculations.
In other cases, strong electron correlations are not adequately described, even by the most advanced $GW$ variants employed in this study.
More often, however, experiments have likely been misinterpreted or were based on suboptimal samples. 

We would like to preface the following considerations with a disclaimer: 
We are not experts on any of the materials discussed. 
On the contrary, we only became aware of them due to their above-average error metrics.
In each case, we can only provide a brief overview of the literature and add our, hopefully sometimes helpful, perspective as computational materials scientists. 
This may inspire new investigations in theoretical and experimental solid state physics, which some of the unusual materials certainly deserve.

For the QP$G_{0}W_{0}$ calculations (lower left panel of Fig.~\ref{fig:histo}), we looked at the outliers AgF and CuLaO$_2$.  
The value of $0.91$~eV for the QP$G_{0}W_{0}$ band gap of AgF severely underestimates the experimental band gap of $2.8 \pm 0.3$~eV \cite{Marchetti1971}.
In contrast, the QS$GW$ band gap of $3.18$~eV agrees well with the experiment.
We attribute this behavior to the metallic ground state predicted by the LDA for AgF. 
Many-body corrections are initially required to open a gap, and self-consistency must be achieved to make that gap substantial.

CuLaO$_2$ is a delafossite oxide that has been studied for its use in photocatalytic hydrogen production \cite{Koriche2006}. 
It should not be confused with La$_2$CuO$_4$, which is well-known for its high-temperature superconductivity and unique electronic properties, cf.~Ref.~\cite{Cunningham2023}.
The CuLaO$_2$ QP$G_{0}W_{0}$ band gap of $4.35$~eV significantly overestimates the experimental band gap of $2.34$~eV \cite{Koriche2006}.
Advancing to QS$GW$ enlarges the band gap to $5.09$~eV, which increases the error and pushes the point off-scale in the QS$GW$ histogram in Fig.~\ref{fig:histo}.
Because a QS$G\hat{W}$ using our benchmark workflow (cf.~Methods) exceeded our memory and compute budget, we recomputed the QS$GW$ and QS$G\hat{W}$ using a reduced basis set and a truncated BSE transition space, see Supplementary~Note~4.
These QS$GW$ and QS$G\hat{W}$ calculations yielded band gaps of $4.93$~eV and $4.36$~eV, respectively.
Although these values are not fully converged, we note that the observed band gap renormalization caused by the inclusion of vertex corrections in $W$ of $0.57$~eV is relatively large.
Based on a limited study of the influence of the BSE transition space on the QS$G\hat{W}$ band gap, we estimate that the converged QS$G\hat{W}$ renormalization is even greater, see Supplementary~Note~4.
However, the resulting band gap still exceeds the experimental value by $2$~eV.
This exactly mirrors the behavior of the related delafossite oxide, CuAlO$_2$, analyzed by Cunningham \textit{et al.}~\cite{Cunningham2023}.
There, the authors demonstrate that the almost dispersionless Cu $3d$ valence band causes a deeply bound exciton, lowering the optical gap to approximately $3.2$~eV, whereas they obtained a fundamental QS$G\hat{W}$ gap of $4$~eV.
The authors further argue that including electron-phonon processes may lower the optical band gap further, as their QS$G\hat{W}$+BSE optical band gap is still larger than the experimental reports for CuAlO$_2$ \cite{Cunningham2023}. 
Additionally, Cunningham \textit{et al.}~\cite{Cunningham2023} noted that the optical absorption edge of CuAlO$_2$ is strongly dependent on preparation and post-annealing conditions, cf.~Ref.~\cite{Li2018}.
Examining the element-resolved QS$GW$ band structure in Supplementary~Note~4 reveals a similar, almost dispersionless valence band composed of Cu states, with minor contributions from O states, for CuLaO$_2$.
Since the experimental band gap of CuLaO$_2$ used in the benchmark comes from a reflectivity measurement, i.e., an optical measurement, we attribute the observed discrepancy to the fact that the measurement most likely report the optical band gap rather than the fundamental electronic band gap.
We also believe that electron-phonon processes and influences from the sample growth process similar to those in CuAlO$_2$ may influence the optical band gap of CuLaO$_2$.
Therefore, further experiments on CuLaO$_2$ as well as a full-scale, in-depth QS$G\hat{W}$+BSE study may be warranted to determine its true optical/fundamental gap.

Among the QS$GW$ outliers (lower middle panel of Fig.~\ref{fig:histo}), we first discuss antifluorite Cu$_2$S and Cu$_2$Se, as both are found to be semimetallic.
In the case of Cu$_2$S we reproduce the results of Lukashev \textit{et al.}~\cite{Lukashev2007}, i.e., that the ideal antifluorite Cu$_2$S remains semimetallic in both LDA and QS$GW$ because of the triple-degenerate Cu $3d$ valence band maximum at the $\Gamma$ point that overlaps the Cu $4s$ conduction band.
Ref.~\cite{Lukashev2007} showed that symmetry breaking through small deviations from the ideal antifluorite positions opens a narrow gap.
However, severe structural disorder, non-stoichiometry, and other effects, as discussed in Ref.~\cite{Lukashev2007}, complicate optical measurements and make calculating the "experimental" gap of $1.1$--$1.2$~eV difficult for Cu$_2$S.

We conjecture that similar physics is at play in Cu$_2$Se.
Refs.~\cite{Rsander2013,Zhang2014,Klan2021} indicate that strong electronic correlations may also play an important role, as expected for partially filled, degenerate $d$ bands.
Zhang \textit{et al.} showed that the treatment of Cu$_2$Se with mBJ$+U$ using a moderate $U=4$~eV for the Cu $d$ states, opens a small band gap of $0.5$~eV, still underestimating the experimental band gap of $1.2$~eV \cite{Madelung2004}.
This result is also consistent with the PBE0$+U$ calculations of Råsander \textit{et al.}~\cite{Rsander2013}, although they used an unusually large value of $U=10$~eV.
A band gap of $0.5$~eV was also obtained though an HSE06$+U$ calculation with $U=4$~eV by Klan~\textit{et al.}~\cite{Klan2021}.
Because the RPA screening used in the QS$GW$ is a poor approximation of short-range correlations \cite{Yan2000}, and DFT$+U$ calculations effectively model short-range, on-site correlations, we presume that short-range correlations are important for obtaining an accurate electronic structure of Cu$_2$Se.
Thus, a more complete picture of the electronic structure of Cu$_2$Se will likely require treating electronic correlations through methods such as QS$GW$+DMFT \cite{Acharya2023}, as well as the inclusion of substantial structural disorder and non-stoichiometry mentioned in Refs.~\cite{Lukashev2007,Zhang2014,Klan2021}. 
We further conjecture that approaches coupling advanced many-body methods with realistic structural models may be required to obtain band gaps that align with experimental results for Cu$_2$S, Cu$_2$Se, and possibly Cu$_2$Te.

In the case of RbAu, a representative of the small and special class of insulating aurides in which Au forms a simple isolable monoatomic anion \cite{Aycibin2014}, we are unsure what exactly causes the deviation of the calculated QS$GW$ band gap of $1.61$~eV to the experimental band gap of $2.75$~eV \cite{Nicoloso1984}.
Interestingly, Ref.~\cite{Nicoloso1984} states that the band gap of RbAu reduced to $1.35 \pm 0.1$~eV upon melting at 510$^\circ$C, which appears to be in better agreement with our calculations. 
However, we did not calculate the liquid state of RbAu.
Earlier work by Liu \cite{Liu1975} describes RbAu as "more like a metallic alloy, with optical properties similar to pure gold in the visible region", referring to a reflectivity feature near $2$~eV.
Based on this limited literature for RbAu, we encourage the experimental community to remeasure the optical response of RbAu under well-controlled conditions and encourage theoretical physicists to look closer into the issue of electronic correlations in aurides. 

Finally, we examine the outliers of our QS$G\hat{W}$ calculations (lower right panel in Fig.~\ref{fig:histo}).
The fact that these are apparently numerous is ultimately also a downside of the success of this method, namely its low standard deviation.
Given the exceptional accuracy and consistency of the QS$G\hat{W}$ method across the benchmark dataset, significant deviations typically suggest either experimental ambiguities or the presence of physics beyond what is captured by the current (purely electronic) many-body treatment.
As the QS$G\hat{W}$ method explicitly includes leading electron-electron vertex corrections in the screened Coulomb interaction $W$, it effectively captures the dominant electronic correlation effects in non-magnetic semiconductors and insulators. 
However, it still omits vertex corrections in $\Sigma$ and does not account for electron-phonon interactions. 
Consequently, pronounced discrepancies might also highlight materials where electron-phonon coupling, polaronic effects, or other subtle many-body phenomena play a crucial role, pointing towards important directions for future theoretical and experimental investigations.

Among the materials for which QS$G\hat{W}$ underestimates the experimental band gaps, the most pronounced outliers are BaTe, BeSe (also highlighted in the QS$GW$ histogram), and BaO.
For BaTe, Martin \textit{et al.}~\cite{Martin1967} measured a band gap of $3.4$~eV using photoelectric emission.
Analyzing the QS$G\hat{W}$ band structure for BaTe, we found an indirect band gap of $2.79$~eV found along the $\Gamma$-X symmetry line and direct band gap of $3.25$~eV at the X point.
Therefore, we argue that Martin \textit{et al.}~\cite{Martin1967} measured the direct band gap at the X point rather than the fundamental indirect gap, explaining the apparent underestimation highlighted in the histogram.
A similar case can be made for the indirect insulator BeSe, where Wilmers \textit{et al.}~\cite{Wilmers1999} report an optical absorption onset E$_0=5.5$~eV derived from ellipsometry data, whereas our QS$G\hat{W}$ calculation predicts an indirect gap of $3.98$~eV along the $\Gamma$-X symmetry line and a direct gap of $6.07$~eV at the $\Gamma$ point.
We therefore argue that Wilmers \textit{et al.}~\cite{Wilmers1999} reported the direct optical band gap at the $\Gamma$ point for BeSe.
A later resonant inelastic X-ray scattering study by Eich \textit{et al.}~\cite{Eich2006} measured a direct band gap of $5.7 \pm 0.2$~eV at the $\Gamma$ point and an indirect band gap of $3.8\pm 0.2$~eV for BeSe.
These values agree much better with our calculated band gaps.
The discrepancy between the values for the direct gap reported by Wilmers \textit{et al.}~\cite{Wilmers1999} and Eich \textit{et al.}~\cite{Eich2006} likely arises from the fact that Wilmers \textit{et al.}~\cite{Wilmers1999} measured the optical gap rather than the fundamental electronic gap.
The main difference between both is the exciton binding energy, estimated here to be around $80$~meV \cite{Grunert2024a}.
The remaining difference is probably due to other experimental uncertainties.

For BaO, our QS$G\hat{W}$ calculation yields a band gap of $4.07$~eV, whereas the experiment by McLeod \textit{et al.}~\cite{McLeod2010} reports a band gap of $4.8$~eV.
However, two older optical measurements report much smaller onset energies for the optical absorption in BaO: $4.10$~eV at room temperature by Saum \textit{et al.}~\cite{Saum1959} and $4.043$~eV for the first bound exciton at $2$~K by Kaneko \textit{et al.}~\cite{Kaneko1988}.
When estimating the exciton binding energy of BaO to be around $80$~meV \cite{Grunert2024a}, our QS$G\hat{W}$ band gap of $4.07$~eV aligns exceptionally well with these older experiments.
Moreover, McLeod \textit{et al.}~\cite{McLeod2010} note a secondary emission band arising from BaCO$_3$ surface carbonation in their spectra, indicating residual CO$_2$ contamination that could have biased their surface-sensitive X-ray emission/absorption spectroscopy band gap determination.
Taken together, these observations call into question the accuracy of the $4.8$~eV band gap reported in Ref.~\cite{McLeod2010} for BaO.
We presume that throughout the dataset, it is not inconceivable that incorrect band gaps may be reported for other materials, especially for those we did not calculate with the QS$G\hat{W}$ method.

The remaining cases in which our QS$G\hat{W}$ calculations underestimate the experimental band gap stem from well-documented shortcomings inherent to the method.
These have been discussed in detail by Kutepov~\cite{Kutepov2022} and by Cunningham \textit{et al.}~\cite{Cunningham2023}.
We provide only a brief and limited summary here and refer the reader to these publications for a more thorough discussion.

First, the QS$G\hat{W}$ method currently omits electron-phonon interactions, which typically reduce calculated band gaps when included \cite{Kutepov2022, Cunningham2023}. 
This omission results in an overestimation of the band gap for lighter compounds where the electron-phonon renormalization (EPR) is strongest.
We discuss this in more detail later.

Secondly, the quasiparticle approximation employed by the QS$G\hat{W}$ approach neglects full self-consistency in Green's function calculations.
Kutepov~\cite{Kutepov2022} analyzed this in detail by starting out from the approximations invoked in \textsc{Questaal} and systematically removing them.
The final conclusion was that additional approximations which are not inherently assumed in Hedin's equations are required to achieve "good results".
Specifically, these approximations include the TDA, the static approximation of the BSE kernel, and excluding vertex corrections in $\Sigma$ \cite{Kutepov2022, Cunningham2023}.
There, it is also noted that omitting vertex corrections in $\Sigma$ helps to avoid the destructive effect of $Z$-factor cancellation (see Appendix~A of Ref.~\cite{Kotani2007}), which occurs when full vertex corrections are used in connection with quasiparticle $GW$ schemes.

A notable example of the shortcomings of the QS$G\hat{W}$ method that is highlighted by both Kutepov~\cite{Kutepov2022} and Cunningham \textit{et al.}~\cite{Cunningham2023} is CuCl, where the inclusion of vertex corrections in $W$ within the quasiparticle self-consistency severely underestimates the band gap.
This is also clearly observed in our benchmark, see Fig.~\ref{fig:histo}.
Our QS$G\hat{W}$ band gap of $2.84$~eV agrees with the values reported by Kutepov~\cite{Kutepov2022} and Cunningham \textit{et al.}~\cite{Cunningham2023}.
Furthermore, in agreement with Refs.~\cite{Kutepov2022, Cunningham2023}, we find that our QS$GW$ band gaps of $3.64$~eV for CuCl and $3.26$~eV for CuBr are closest to the experimental values of $3.4$~eV and $3.07$~eV \cite{Madelung2004}, respectively.
Cunningham \textit{et al.}~\cite{Cunningham2023} point out that such underestimations generally occur when the highest occupied bands are flat and almost dispersionless, matching our observations for CuBr, and the rare gas solids, Ar and Kr.
Furthermore, we observe that these underestimations are most prominent in materials where EPR effects are negligible, as no error cancellation occurs.
By "error cancellation", we mean that the missing EPR reduction of the band gap is compensated by the systematic underestimation of the band gap in the QS$G\hat{W}$.
Thus, in a somewhat ironic way, the accuracy of the QS$G\hat{W}$ seems to rely on fortuitous error cancellation, somewhat similar to $G_{0}W_{0}$ calculations.
Diamond is a prime example of this, with an EPR of about $0.4$~eV \cite{Cardona2005, Giustino2010, Monserrat2014}.
There, the QS$G\hat{W}$ band gap of $5.64$~eV agrees with the measured band gap of $5.5$~eV \cite{Madelung2004} only because the systematic band gap underestimation from the $W$-only vertex correction nearly cancels out the missing EPR gap reduction.
This observation has also been made previously by Kutepov~\cite{Kutepov2022}.
We would like to emphasize that addressing and ameliorating these systematic shortcomings continues to be an active area of ongoing research.

We turn now to the cases where QS$G\hat{W}$ overestimates the experiment, focusing first on LiF and solid Ne.
As previously mentioned, neglecting EPR in these light-element, wide-gap insulators results in a significant overestimation of the band gaps, particularly when unfavorable error cancellation occurs.
By contrast, the discrepancy between the published experimental values for MgCl$_2$ \cite{Thomas1973}, PbF$_2$ \cite{Findley1983} and SrF$_2$ \cite{Taylor1988}, and our QS$G\hat{W}$ calculations, are presumably caused again by the fact that the measurements report optical gaps (i.e., optical absorption onsets) rather than fundamental electronic band gaps.
We were able to confirm this for SrF$_2$ using an older measurement by Rubloff~\cite{Rubloff1972}, but we could not find comparable data for the fundamental gap of MgCl$_2$ and PbF$_2$.
Since these wide-gap fluorides have weak dielectric screening and thus large exciton binding energies (possibly up to $1$~eV) \cite{Grunert2024a}, their absorption onsets are well below the conduction band edge. 
We again presume that optical band gaps have been reported for other materials in the benchmark dataset.
However, these are difficult to find unless the experiments are reexamined or a QS$G\hat{W}$ calculation is performed for them.

Since the aforementioned misreported band gaps (direct vs. indirect and optical vs. fundamental) and materials, such as Cu$_2$S and Cu$_2$Se, for which the experimental crystal structures probably differ from the high-symmetry structure used in the calculations, bias the error statistics, we have recalculated all metrics for a sanitized dataset, see Supplementary~Note~5.
In addition to excluding the eight previously mentioned materials, i.e., CuLaO$_2$, Cu$_2$S, Cu$_2$Se, MgCl$_2$, PbF$_2$, SrF$_2$, RbAu, and BaTe, we also adjusted the reference experimental band gaps for BeSe and BaO (see above), as well as ScN.
Regarding ScN, we observed that Cunningham \textit{et al.}~\cite{Cunningham2023} also encountered difficulty in explaining the discrepancy between their QS$G\hat{W}$ of $1.27$~eV (here: $1.29$~eV) and the experimental values of $0.9$~eV by Al-Brithen \textit{et al.}~\cite{Brithen2004} and $0.92$~eV by Deng \textit{et al.}~\cite{Deng2015}, even when approximating the band gap reduction caused by EPR to be around $0.1$~eV.
However, a more recent work by Grümbel \textit{et al.}~\cite{Gruemble2024}, presumably using a cleaner sample, reported a band gap of $1.1\pm 0.1$~eV, aligning more closely with the results of Cunningham \textit{et al.}~\cite{Cunningham2023} and this study.

To make the results comparable to those of previous publications that also used the Borlido \textit{et al.}~\cite{Borlido2019, Borlido2020} benchmark dataset, the main text still reports results on the original dataset.
However, we recommend using the corrected benchmark (provided as additional Supplementary Material available online), which omits or reclassifies materials with incorrectly reported band gaps.
We also strongly believe that the original benchmark dataset contains more materials with misreported band gaps and structures, particularly for materials with larger unit cells for which we did not perform any $GW$ calculation.

\section*{Discussion}

We conducted a systematic benchmark for band gaps of solids, comparing MBPT methods, specifically, various $GW$ approximations, with two state-of-the-art DFT functionals, namely mBJ and HSE06, on a large dataset of non-magnetic semiconductors and insulators. 
Addressing the central research question raised in this study---\textit{"How well do MBPT calculations perform relative to the best available DFT functionals when predicting the band gaps of solids?"}---our findings clearly demonstrate the superior performance of advanced $GW$ schemes.
Specifically, we found that $GW$ methods such as the QP$G_{0}W_{0}$ and the QS$G\hat{W}$ significantly outperform the current best DFT functional (mBJ and HSE06) and simpler $GW$ variants ($G_{0}W_{0}$-PPA). 
As expected, the QS$G\hat{W}$ emerged as the most accurate method overall.

Interestingly, although the computationally cheaper $G_{0}W_{0}$ approaches exhibited good average accuracy, their reliability heavily depends on fortuitous error cancellations. 
These cancellations arise due to balancing opposite systematic errors, i.e., overestimated dielectric screening from the DFT starting point and underestimated dielectric screening due to the use of the RPA and the absence of excitonic effects (vertex corrections in $W$). 
Our benchmark again highlights the sensitivity of $G_{0}W_{0}$ results to the choice of the DFT functional used as a starting point.
Notably, contrary to common practice, performing a $G_{0}W_{0}$ calculation starting from an LDA DFT calculation performs better on average than those starting from a PBE DFT calculation.

Furthermore, the observed performance of $G_{0}W_{0}$-PPA calculations compared to DFT calculations depends heavily on whether the full dataset or a subset of materials with complete data coverage is considered.
We argue that complete data coverage subset provides a fairer basis for comparison, and on this basis, $G_{0}W_{0}$-PPA slightly outperforms the meta-GGA functional mBJ and notably surpasses the hybrid functional HSE06. 
Nonetheless, both DFT functionals remain substantially more computationally efficient and easier to converge (cf.~Ref.~\cite{Grossmann2024}). 
Particularly, the excellent performance combined with computational affordability explains the widespread adoption of HSE06, especially for large-scale applications involving, e.g., surfaces \cite{Großmann2025} and defects \cite{Flötotto2025}. 
However, caution should be exercised  when using the mBJ functional, given its semi-empirical nature, as it was specifically fitted to experimental band gaps \cite{Tran2009}. 
Moreover, mBJ and other meta-GGA functionals often exhibit stability issues, convergence difficulties, and generally struggle to provide accurate electronic densities, exchange-correlation energies, and band gaps simultaneously \cite{Aouina2024}.

Even though the DFT functionals used for comparison in this benchmark are surprisingly good and well-performing, our findings show that MBPT in the form of different $GW$ variants can contribute a lot to our understanding and that ultimately at least the QS$G\hat{W}$ methods simply has the superior predictive power.
Note that the QP$G_{0}W_{0}$ is an attractive alternative to the QS$G\hat{W}$ due to the intrinsic error compensation discussed above and the lower costs.

Our results also underscore the importance of incorporating spin-orbit coupling (SOC) into $GW$ calculations for heavy-element materials. 
SOC significantly improves the accuracy of predictions for materials such as PbTe and SnTe, demonstrating its essential role in accurately capturing the nuances of the electronic structure of heavy-element semiconductors.

Despite recent methodological advances, limitations still remain in current MBPT approaches.
For instance, despite its superior accuracy, the QS$G\hat{W}$ approach systematically underestimates the band gaps of materials with flat or almost dispersionless highest occupied bands, and of materials for which EPR effects are negligible.
This systematic underestimation stems from the approximations inherent in the quasiparticle self-consistent method and the QS$G\hat{W}$ itself, including the TDA, the static BSE kernel, and the exclusion of vertex corrections in $\Sigma$. 

However, the accuracy of the QS$G\hat{W}$ method is arguably already within the experimental tolerance for most materials, which underscores its strength as a high-fidelity predictive tool. 
Our detailed examination of outliers also revealed several experimental ambiguities, such as confusion between optical and fundamental band gaps, as well as between direct and indirect transitions and the influence of structural anomalies.

Additionally, omitting electron-phonon interactions from MBPT calculations is a significant limitation, especially for light elements and wide-gap insulators. 
Systematically incorporating these interactions would enhance the predictive power of MBPT calculations, especially for materials with strong electron-phonon coupling.

Finally, our benchmark also provides critical guidance for constructing reliable datasets for machine learning applications in materials science from a broader perspective. 
Despite the high computational cost of MBPT calculations, the strategic use of MBPT methods to generate small, high-quality datasets for transfer learning seems promising. 
This approach could significantly improve the reliability and accuracy of future machine-learning-assisted materials predictions while balancing computational affordability with the necessity of precision in materials discovery.

\section*{Methods}

All calculations described below use the same experimental geometries from the ICSD \cite{Bergerhoff1983, Belsky2002, Zagorac2019} that were used in the DFT benchmark by Borlido \textit{et al}.~\cite{Borlido2019, Borlido2020}.
The corresponding material identifiers can be found either in the DFT benchmark dataset by Borlido \textit{et al}.~\cite{Borlido2019, Borlido2020} or in the provided Supplementary Material.
All structures were reduced to their primitive standard structure using \textsc{Spglib} \cite{spglib}.
For all k‑point grids (i.e., $k_x \times k_y \times k_z$) used in calculations described below, each subdivision along the $x$, $y$, and $z$ directions (i.e., $k_x$, $k_y$, and $k_z$) was rounded up to the nearest even integer.
We provide a great deal of detail in the Methods section because the value of any benchmark depends on its reproducibility and its character as "good practice" template for own work.

\subsection*{$G_{0}W_{0}$-PPA using Quantum ESPRESSO and Yambo}

We performed DFT calculations with the plane-wave code \textsc{Quantum ESPRESSO} (version 7.1) \cite{Giannozzi2009all, Giannozzi2017all} as starting point for the $G_{0}W_{0}$-PPA calculations, using the LDA and PBE exchange-correlation functionals \cite{Marques2012}.
For the LDA calculations, we used optimized norm-conserving Vanderbilt (ONCV) pseudopotentials from the PseudoDojo project (version 0.4.1) \cite{Hamann2013, vanSetten2018}.
For the PBE calculations, we used ONCV pseudopotentials from the SG15 library (version 1.2) \cite{Hamann2013}.

The $G_{0}W_{0}$-PPA calculations were carried out using the \textsc{Yambo} code (version 5.2.4).
The frequency dependence of the dynamical screening $W$ was approximated through the Godby-Needs plasmon-pole approximation \cite{Godby1989}. 
To accelerate the convergence of $\Sigma$ with respect to the number of empty bands, $N_\mathrm{b}$, the Bruneval-Gonze technique was employed \cite{Bruneval2008}.
The $\mathbf{q}\rightarrow 0$ divergence of the Coulomb potential was treated with the random integration method \cite{Marini2009} as implemented in \textsc{Yambo}. 
The same plane-wave cutoff energy was used in \textsc{Yambo} to expand the KS wave functions and densities in plane waves as was used for the converged DFT calculation in \textsc{Quantum ESPRESSO}.

For the $G_{0}W_{0}$@LDA-PPA and $G_{0}W_{0}$@PBE-PPA calculations, we used the following workflow: 

(i) As a first step, we converged the DFT plane-wave cutoff energy and k-point grid.
The starting value for the plane-wave cutoff energy was set to the value suggested by the respective pseudopotentials and for the k-point grid the initial value was set to a structure-independent reciprocal density of $n_\mathbf{k} = 1500$ per atom as defined in \textsc{pymatgen} \cite{Ong2013}.
We then iteratively increase the plane-wave cutoff energy by $5$~Ry, while keeping $n_\mathbf{k}$ fixed, until the change in the total energy per atom is smaller than $1$~meV. 
Then, using the found plane-wave cutoff energy, we converged the total energy change per atom with respect to the k-point grid to the same $1$~meV threshold.
For this, we increased the k-point grid, $k_x \times k_y \times k_z$, in the following manner: 
We increased the k-point grid density, $n_\mathbf{k}$, until the number of subdivisions in some direction of the resulting new k-point grid was greater than that of the previous k-point grid, i.e., $k_i^{n+1} > k^n_i$ for some $i = \{x, y, z\}$.
Then, we performed a DFT calculation using the new, denser k-point grid and compared the total energy per atom to the previous one.
This process was repeated until the total energy change per atom converged below $1$~meV.

(ii) Similar to Borlido \textit{et al}.~\cite{Borlido2019, Borlido2020}, we converge the band gap with respect to the k-point grid, by increasing the k-point grid from the previously found $n_\mathbf{k}$ until the changes in the band gap are below a threshold of $25$~meV.
Here, the plane-wave cutoff energy is set to the converged value from step (i).

(iii) Starting from converged KS wavefunctions and eigenvalues from step (i), we converged the $G_{0}W_{0}$-PPA corrections with respect to the number of empty bands, $N_\mathrm{b}$, the plane-wave cutoff energy used for the dielectric matrix, $G_\mathrm{cut}$, and the $GW$ k-point grid density $n_\mathbf{k}^\mathrm{GW}$.
These three parameters were converged using a workflow previously introduced by us \cite{Grossmann2024}.
The starting points and step sizes of the parameters were set to the values from Ref.~\cite{Grossmann2024}.
The convergence threshold for the direct band gap at the $\Gamma$ point was set to $25$~meV.

(iv) To ensure the highest possible accuracy at a moderate increase in computational cost for the $G_{0}W_{0}$-PPA calculations, we increased the parameters found in (iii) (i.e, $N_\mathrm{b}$, $G_\mathrm{cut}$, and $n_\mathbf{k}^\mathrm{GW}$) by one additional step and performed a final $G_{0}W_{0}$-PPA calculation.
To save computational resources, we only calculate the $G_{0}W_{0}$-PPA band gap correction at the $\Gamma$ point for the valence band maximum and the conduction band minimum.

(v) The band gap correction obtained from the $G_{0}W_{0}$-PPA calculation at the $\Gamma$ point was then used as a "scissor operator", i.e., the obtained value was added to the converged DFT gaps from step (ii), resulting in the band gaps reported in this study.

\subsection*{All-electron $GW$ calculations in Questaal}

We performed all-electron DFT and QP$G_{0}W_{0}$, QS$GW$ and QS$G\hat{W}$ calculations in the LMTO code \textsc{Questaal} (version 7.14.1).
For details regarding the all-electron LMTO implementation, QS$GW$ methodology and how vertex corrections are included in $W$, we refer to the work by Pashov \textit{et al.}~\cite{Pashov2020}, Kotani \textit{et al.}~\cite{Kotani2007}, and Cunningham \textit{et al.}~\cite{Cunningham2023}, respectively.

A particularly useful feature of the \textsc{Questaal} code is its ability to interpolate $\Sigma_0$, see Eq.~(\ref{eq:qp}), onto an arbitrary k-point grid \cite{Pashov2020}.
To take advantage of this feature, $GW$ calculations in \textsc{Questaal} typically use a different k-point grid for calculating $\Sigma$ than for solving the KS equations (i.e., performing a DFT calculation).
From now on, we will refer to the DFT k-point grid as $k_\mathrm{DFT}$ and the $\Sigma$ k-point grid as $k_{\Sigma}$.
Since it is empirically known that $\Sigma$ is more local and therefore tends to converge much faster with respect to the k-point grid than a DFT calculation \cite{Pashov2020, Cunningham2023, Grossmann2024}, it is highly efficient to calculate $\Sigma$ on a relatively coarse k-point grid, and then solve the effective KS equations, in which $V_{xc}^\text{KS}$ is replaced by $\Sigma_0$ interpolated on a fine k-point grid.
First, this allows us to decouple the convergence of $k_\mathrm{DFT}$ and $k_{\Sigma}$.
Second, $GW$ band structures and other post-processing steps, such as a density of states (DOS) calculation, can be performed essentially at the cost of a DFT calculation, after $\Sigma$ is obtained.

We begin by describing the general calculation setup:
Using symmetrized structures as described above, we employed the automatic input generator provided with \textsc{Questaal}, \texttt{blm}, to generate input files for the crystal structures in the same manner as Cunningham \textit{et al.}~\cite{Cunningham2023}.
Unless a parameter is explicitly mentioned below, all parameters were used at their default values generated by \texttt{blm}, which are \textsc{Questaal} defaults in most cases.
We generally enable \texttt{blm}'s  \texttt{gw} flag, generating a larger basis set, better suited for accurate $GW$ calculations.
In line with the \textsc{DeltaCodes} project \cite{Lejaeghere2016}, we adjusted the criteria to include more semicore states than the default settings within the basis set to increase the accuracy of the results.
Specifically, we included an atomic state as an extended local orbital (ELOs)---local orbitals with a smooth Hankel function tail that extends continuously and differentiable into the interstitial region \cite{Pashov2020}---in the basis if either the atomic state lies higher than $\texttt{eloc}=-2.5$~Ry or the share of its corresponding charge beyond the LMTO augmentation sphere is greater than $\texttt{qloc}=0.002$.
In addition, we included "floating orbitals" \cite{Pashov2020} to improve the completeness of the basis set in interstitial regions.
These were generated and placed automatically by \texttt{blm}.
\textsc{Questaal} represents the interstitial part of the basis functions with plane waves \cite{Pashov2020}, thereby introducing an energy cutoff parameter analogous to the plane-wave cutoff used in \textsc{Quantum ESPRESSO}.
At this point, it is important to note that DFT and $GW$ calculations in \textsc{Questaal} use entirely separate codes and, therefore, different plane-wave cutoffs.
The plane-wave cutoff parameter \texttt{gmax} used for the DFT calculations is provided by \textsc{Questaal}'s \texttt{lmfa} code, which is also responsible for building the basis set and computing the free-atom charge densities.
It provides two cutoff values by default: one sufficient for accurately describing the valence states and another, larger one that accurately describes the deep core-like states included in the basis as well.
This cutoff choice is the all-electron equivalent of the so-called "hardness" used to characterize pseudopotentials in plane-wave pseudopotential codes.
We always used the higher value for the benchmark workflow.
The $GW$ code of \textsc{Questaal} differentiates between one- and two-particle basis sets and requires setting two additional cutoff energies.
Here, the two-particle basis set is an auxiliary basis set of wave function products required for $GW$ calculations.
The cutoff energies for both basis sets were generated using an internal algorithm included in \texttt{blm}, which sets these cutoffs very conservatively.
In other words, it sets them much larger than necessary.
To verify the cutoffs used in the $GW$ code of \textsc{Questaal} for our workflow, we compared them to those used by Cunningham \textit{et al.}~\cite{Cunningham2023} for some systems.
We found that our cutoff values ($G_\mathrm{cut}(\Psi, M)$ in Ref.~\cite{Cunningham2023}) are consistently higher than the values used there.
At last, we adjusted the parameter that controls the previously mentioned interpolation of $\Sigma_0$. 
Since the high-energy parts of $\Sigma_0$ do not interpolate well \cite{Pashov2020}, \textsc{Questaal} sets them to zero above a cutoff energy $\Sigma_\mathrm{cut}$.
As Cunningham \textit{et al.}~\cite{Cunningham2023} have already noted, the results weakly depend on $\Sigma_\mathrm{cut}$.
To be safe, we increased the cutoff energy from the default of $2$~Ry, to $3$~Ry (cf.~Ref.~\cite{Cunningham2023}, Tab.~XII).

Although automatically generated basis sets are convenient, they can develop pathologies that cause calculations to become unstable.
For this reason, we had to manually intervene in our benchmark for the compounds containing Hf, Ta, W, and Tl.
For Hf, Ta, and W, we exclude the high-lying $d$ orbitals because they interfered with the low-lying $4f$ orbitals.
In turn, we added $5p$ orbitals as ELOs to the basis set.
For Ta and W, we also added $4f$ orbitals as ELOs.
In the case of Hf, we added $5f$ orbitals as a high-lying local orbital everywhere except for HfSe$_2$, where they caused instabilities in the calculations. 
For Tl, all local orbitals except the $5d$ ELOs were removed.
We limited the maximum angular momentum of orbitals inside the muffin-tin spheres to $\texttt{lmx} = 4$ for all four elements, and we set the angular-momentum cutoff for projecting wave-function tails centered on neighboring sites to $\texttt{lmxa} = 5$ for Ta, W and Tl and $\texttt{lmxa} = 6$ for Hf.
In addition to these manual adjustments to the basis set for these four elements, we had to set \texttt{qloc} back to its default value of $0.005$ for AlCuO$_2$, GaCuO$_2$, LaCuO$_2$, ScCuO$_2$, and Cu$_2$O in order to obtain stable calculations.

Next, we describe the computational workflow for performing and converging QP$G_{0}W_{0}$, QS$GW$ and QS$G\hat{W}$ calculations automatically using the input settings and basis set described above:

(i) As a first step, we converged the DFT k-point grid $k_\mathrm{DFT}$.
Similar to our \textsc{Quantum ESPRESSO} workflow, the initial value for the DFT k-point grid was set to a structure-independent reciprocal density of $n_\mathbf{k} = 1000$ per atom as defined in \textsc{pymatgen} \cite{Ong2013}.
(For CsAu, BaO, and anatase TiO$_2$, we set the initial density of the DFT k-point grid to $n_\mathbf{k} = 1500$.)
Then, we converged the total energy change per atom with respect to the k-point grid to less than $10^{-4}$~Ry, by iteratively increasing the k-point grid in the same manner as described above for step (i) of the \textsc{Quantum ESPRESSO} and \textsc{Yambo} $G_{0}W_{0}$-PPA workflow.
To reduce the computational effort, we performed one iteration of the KS self-consistency cycle per DFT calculation and used the Harris-Foulkes energy \cite{Foulkes1989} per atom as our "total energy" for the k-point grid convergence.
We then performed fully self-consistent DFT using the found converged k-point grid, i.e., $k_\text{DFT}$.

(ii) In the next step, we converged the self energy k-point grid $k_\Sigma$ using QP$G_{0}W_{0}$ calculations by increasing $k_\Sigma$ in the same manner as $k_\mathrm{DFT}$.
However, the initial value of the density of the $\Sigma$ k-point grid was lowered to $n_\mathbf{k} = 300$, as $\Sigma$ tends to converge much faster with respect to the k-point grid than a DFT calculation \cite{Pashov2020, Cunningham2023, Grossmann2024}.
(For CsAu, BaO, and anatase TiO$_2$, we set the initial density of the $\Sigma$ k-point grid to $n_\mathbf{k} = 1000$.)
We increased $k_\mathrm{\Sigma}$ until the difference in QP$G_{0}W_{0}$ band gaps was less than $25$~meV or $k_\mathrm{DFT}$ was reached.
In rare cases where convergence to below $100$~meV was not achieved when $k_\mathrm{DFT}$ was reached, we omitted the QP$G_{0}W_{0}$ and the results of the subsequent calculations for the materials in question from the benchmark dataset.
We found that half of the DFT k-point grid density or less was in most cases sufficient for convergence, again highlighting the more local character of the self energy.
To obtain the QP$G_{0}W_{0}$+SOC band gap, we performed a post-processing step in which we solved the effective KS equations, including the interpolated $\Sigma_0$ and the diagonal parts of $\mathbf{L}\cdot \mathbf{S}$, as described in detail by Pashov \textit{et al.}~\cite{Pashov2020} (cf.~Section~2.8.2 and Section~3.9).

(iii) Because the subsequent QS$GW$ and QS$G\hat{W}$ steps are much more demanding, we use the penultimate $k_\Sigma$ from step (ii) for evaluating $\Sigma$ from this step onward.
This reduces wall time by nearly an order of magnitude while maintaining the band gap difference within our convergence criterion of 25 meV (or 100 meV in rare cases, see above).
Starting from the QP$G_{0}W_{0}$ self energy, we iterated the QS$GW$ cycle until the root mean square (RMS) change in the static part of $\Sigma_0$ reaches $10^{-5}$~Ry.
We stopped the self-consistency cycle if a calculation did not converge by the $25$th iteration.
This was done because we noticed that when convergence was difficult to achieve, the RMS tended to oscillate around and below $10^{-4}$.
However, the significant digits of the band gap did not change during these RMS oscillations.
Fortunately, this has only rarely happened.
Also, note that stopping after iteration $25$ if convergence is not achieved by then also saves computing resources.

(iv) Starting from the QS$GW$ self energy, we iterated the QS$G\hat{W}$ in the same manner as in step (iii), again using the penultimate $k_\Sigma$ from step (ii).
The transition space, i.e., the number of valence and conduction bands included in the Bethe-Salpeter Hamiltonian, was set using a heuristic, as converging QS$G\hat{W}$ calculations for all materials with respect to the BSE transition space is computationally infeasible:
We simply included in the Bethe-Salpeter Hamiltonian all bands $10$~eV below (above) the valence band maximum (conduction band minimum).
Additionally, we include bands more than $10$~eV below the valence band maximum as long as each subsequent band is separated from the previous one by less than $0.5$~eV.
This is done to include all bands from a specific valence band manifold, e.g., some semicore states, in the BSE transition space.
For materials that were also calculated by Cunningham \textit{et al.}~\cite{Cunningham2023}, we compared the obtained BSE transition spaces and found that our heuristic produced transition spaces that were equal to or larger than those used there.
Similar to step (iii), the self-consistency cycle did not converge after $25$ iterations for a few materials.
We observed again that the significant digits of the band gap did not change anymore, and the RMS fluctuated around and below $10^{-4}$.

As a post-processing step after each workflow step, we calculated band structures (including element projections) and DOS for each calculated theory level, except for calculations that include SOC.
All band structures were calculated based on k-point paths taken from Ref.~\cite{Hinuma2017} as implemented in \textsc{Questaal}.
These can be found in the provided database entries, see Data and Code Availability statement.

To extract band gaps from our all‑electron \textsc{Questaal} calculations, we combined two Brillouin zone sampling strategies---(i) a regular three-dimensional k-point grid ($E_\mathrm{gap}^{\mathbf{k}_\mathrm{DFT}}$); (ii) k-point paths along symmetry lines \cite{Hinuma2017} ($E_\mathrm{gap}^{\mathbf{k}_\mathrm{sym}}$), i.e., a band structure calculation---and, for each system, reported:
\begin{equation*}
    E_\mathrm{gap} = \min\left(E_\mathrm{gap}^{\mathbf{k}_\mathrm{DFT}}, E_\mathrm{gap}^{\mathbf{k}_\mathrm{sym}} \right)
\end{equation*}
For materials whose extrema lie along high-symmetry lines---silicon being the canonical example---a conventional band structure path correctly determines the band gap.
In contrast, for materials such as LiCoO$_2$ \cite{Radha2021}, the band extrema can be located at general k-points anywhere in the Brillouin zone.
Thus, a band structure calculation cannot capture them, and a regular three-dimensional k-point grid is required.
However, the average (signed) difference between the two approaches is in our case only about $4$~meV, with a mean absolute difference of about $7$~meV. 
Outliers exist, though, such the aforementioned LiCoO$_2$ \cite{Radha2021}, see Supplementary~Note~6.
Since we did not compute band structures for the QP$G_{0}W_{0}$+SOC and QP$G\hat{W}$+SOC calculations, we omit band gaps for these methods from the benchmark whenever the LDA $E_\mathrm{gap}^{\mathbf{k}_\mathrm{sym}}$ was more than $25$~meV smaller than $E_\mathrm{gap}^{\mathbf{k}_\mathrm{DFT}}$, which occurred rarely.

In conclusion, the automated execution of the complex workflow outlined above---which involves performing and converging all-electron QP$G_{0}W_{0}$ followed by self-consistent QS$GW$ and its vertex-corrected variant, QS$G\hat{W}$---for a large set of materials is a formidable computational challenge.
As described above, we acknowledge that a few pragmatic shortcuts were unavoidable.
Nevertheless, we are confident that the reported band gaps are converged to an accuracy of about $100$~meV or better.

\section*{Data and Code Availability}

The high-throughput MBPT workflows used in this study, along with the generated data and post-processing code, are available at \url{https://github.com/MaxGrossmann/mbpt_benchmark} or \url{https://doi.org/10.5281/zenodo.16759144}.

\section*{Acknowledgments}

We thank the staff of the Compute Center of the Technische Universität Ilmenau and especially Mr.~Henning~Schwanbeck for providing an excellent research environment.
The authors thank W.~G.~Schmidt (Paderborn, Germany) for valuable advice and support regarding computations at the Paderborn Center for Parallel Computing (PC2).
We would also like to thank J.~Jackson (Daresbury, United Kingdom) and B.~Cunningham (Belfast, United Kingdom) or their insightful advice on performing and converging QS$GW$ and QS$G\hat{W}$ calculations using the \textsc{Questaal} code.
Sincere thanks also go to M.~A.~L.~Marques (Bochum, Germany) and J.~Schmidt (Zürich, Switzerland) for inspiring discussions.
This work is supported by the Deutsche Forschungsgemeinschaft DFG (project 537033066) and the Carl Zeiss Stiftung (funding code: P2023-02-008).

\section*{Competing interests}

The authors declare no competing interests.

\section*{Author contributions}

M.G.~and M.G.~conceived the idea; M.T.~wrote the workflow and performed all the computations using the \textsc{Quantum ESPRESSO} and \textsc{Yambo} codes, M.~Großmann wrote the workflow and carried out all calculations using the \textsc{Questaal} code; M.~Grunert optimized the LMTO basis sets for materials containing Hf, Ta, W, and Tl; M.T.~and M.~Großmann analyzed the data and visualized the results; M.~Großmann wrote the first draft of the manuscript; E.R.~supervised the work; all authors revised and approved the manuscript.

\bibliography{literature}

\end{document}


\title{Supplementary Information for "Many-body perturbation theory vs. density functional theory: A systematic benchmark for band gaps of solids"}

\author{Max Großmann}
\email{max.grossmann@tu-ilmenau.de}
\affiliation{Institute of Physics and Institute of Micro- and Nanotechnologies, Technische Universit\"at Ilmenau, 98693 Ilmenau, Germany}
\affiliation{These authors contributed equally to this work.}

\author{Marc Thieme}
\affiliation{Institute of Physics and Institute of Micro- and Nanotechnologies, Technische Universit\"at Ilmenau, 98693 Ilmenau, Germany}
\affiliation{These authors contributed equally to this work.}

\author{Malte Grunert}
\affiliation{Institute of Physics and Institute of Micro- and Nanotechnologies, Technische Universit\"at Ilmenau, 98693 Ilmenau, Germany}

\author{Erich Runge}
\affiliation{Institute of Physics and Institute of Micro- and Nanotechnologies, Technische Universit\"at Ilmenau, 98693 Ilmenau, Germany}

\date{\today}

\maketitle

For more information on abbreviations, please refer to the main text, where all abbreviations are defined in detail.
Any abbreviations not used in the main text will be defined here.

\tableofcontents
\newpage
\clearpage

\section*{Supplementary Note 1: PAW vs.~NC vs.~AE}

The starting point for any $GW$ calculation of our MBPT benchmark was an LDA or PBE DFT calculation.
Since the DFT calculations were performed using different codes and formulations than in the original Borlido \textit{et al.}~\cite{Borlido2019, Borlido2020} benchmark, comparing the DFT band gaps is sensible.
In Supplementary~Fig.~\ref{fig:start_cmp_lda}, we compare the LDA band gaps obtained from \textsc{VASP} \cite{Kresse1996, Kresse1999} with projector-augmented wave (PAW) pseudopotentials \cite{Bloechl1994} by Borlido \textit{et al.}~\cite{Borlido2019, Borlido2020} with those obtained from our \textsc{Quantum ESPRESSO} \cite{Giannozzi2009all, Giannozzi2017all} calculations with norm-conserving (NC) pseudopotentials \cite{Hamann2013, vanSetten2018} and our LMTO all-electron (AE) calculations using the \textsc{Questaal} code \cite{Pashov2020}.
The data cluster tightly around the angle bisector through the origin ($y\!=\! x$) in the top row of Supplementary~Fig.~\ref{fig:start_cmp_lda}, confirming the overall consistency of the three \textit{ab initio} codes.
However, a closer look at the low-gap regime (bottom row) reveals some discrepancies.
Some compounds that are metallic in NC LDA or AE LDA calculations appear as narrow-gap semiconductors in PAW LDA calculations, details in Supplementary~Tab.~\ref{tab:start_cmp_lda}.

\begin{figure}[ht]
    \centering
    \includegraphics{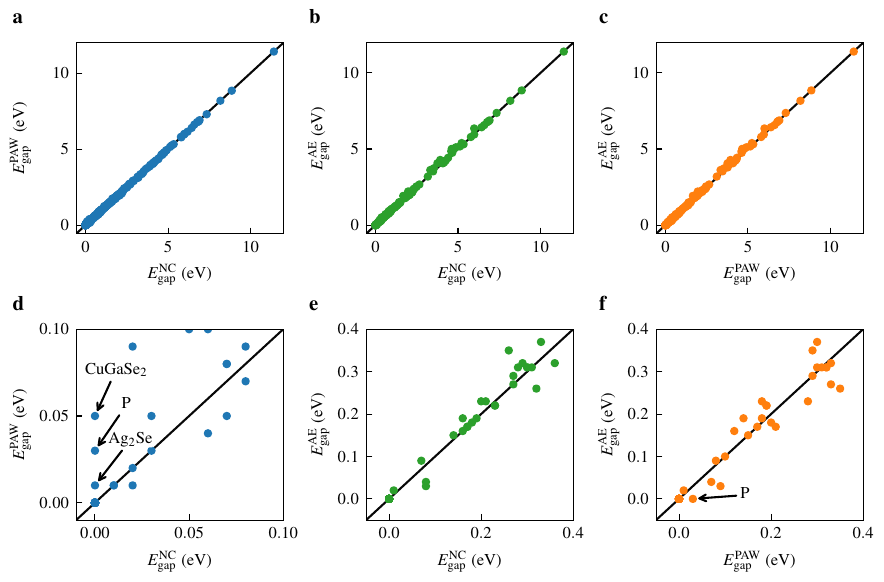}
    \caption{
    \textbf{Comparison of LDA band gaps obtained with PAW, NC, and AE calculations.}
    The top row (i.e., \textbf{a}, \textbf{b}, and \textbf{c}) shows the full band gap range.
    The bottom row (i.e., \textbf{d}, \textbf{e}, and \textbf{f}) zooms into the low band gap region. 
    The solid black line highlights $y\!=\!x$.
    }
    \label{fig:start_cmp_lda}
\end{figure}

\begin{table}[ht]
    \centering
    \caption{
    \textbf{Materials for which LDA PAW predicts a narrow band gap, but LDA NC and/or LDA AE predict a metallic state.}
    Band gaps are given in eV.
    }
    \begin{threeparttable}
        \begin{tabular}{ccccc}
        \toprule
        Composition ~&~ ICSD ID ~&~ LDA PAW (eV) ~&~ LDA NC (eV) ~&~ LDA AE (eV) \\
        \midrule
        Ag$_{2}$Se & 260148 & 0.01 & 0.00 & --- \\
        CuGaSe$_{2}$ & 42097 & 0.05 & 0.00 & --- \\
        P & 417180 & 0.03 & 0.00 & 0.00 \\
        \bottomrule
        \end{tabular}
    \end{threeparttable}
    \label{tab:start_cmp_lda}
\end{table}

Given that we converged the band gaps with respect to the k-point grid for the NC DFT calculations, analogous to Borlido \textit{et al.}~\cite{Borlido2019, Borlido2020} (see the Methods section in the main text), and obtained the AE LDA band gap by combining two Brillouin zone sampling strategies (see Supplementary~Note~6), we are confident that this observation is not due to convergence issues.
Therefore, we suspect that the discrepancy is due to the difference in valence states included in the PAW and NC pseudopotentials for Cu, Ag, and Se.
Compared to the PAW pseudopotentials, the NC pseudopotentials include the $3s$ and $3p$ states as semicore states for Cu, the $4s$ and $4p$ states for Ag, and the $3d$ states for Se.
Therefore, it seems that for Ag$_2$Se and CuGaSe$_2$ the addition of deeper semicore states to the valence band is important for correctly identifying the correct LDA ground state, as the NC calculations agree well with the AE calculations.
However, for P (black phosphorus), there is no difference in the valence configuration between the PAW and NC pseudopotentials.
We conjecture that the missing norm-conservation condition of the PAW may cause this difference, but as emphasized in the main text: Any outlier may indicate an interesting case, worthwhile to be studied in more detail. 
The difference in plane-wave cutoffs may also be responsible for the discrepancy, as PAW pseudopotentials tend to be "softer", that is, they require a smaller cutoff than NC pseudopotentials.

\begin{figure}[ht]
    \centering
    \includegraphics{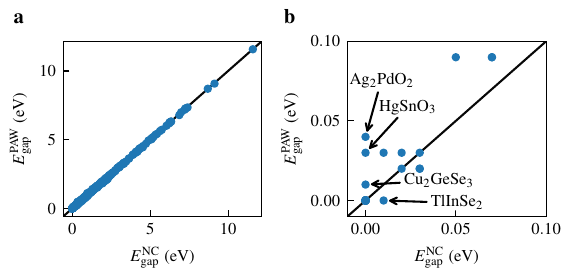}
    \caption{
    \textbf{Comparison of PBE band gaps obtained with PAW and NC DFT calculations.}
    The left panel \textbf{a} shows the full band gap range.
    The right panel \textbf{b} zooms into the low band gap region. 
    The solid black line highlights $y\!=\!x$.
    }
    \label{fig:start_cmp_pbe}
\end{figure}

\begin{table}[ht]
    \caption{
    \textbf{Materials for which PAW PBE predicts a narrow band gap (metallic state), but NC PBE predict a metallic state (narrow band gap).}
    Band gaps are given in eV.
    }
    \centering
    \begin{threeparttable}
        \begin{tabular}{cccc}
        \toprule
        Composition ~&~ ICSD ID ~&~ PBE NC (eV) ~&~ PBE PAW (eV) \\
        \midrule
        Ag$_{2}$PdO$_{2}$ & 51498 & 0.00 & 0.04 \\
        Cu$_{2}$GeSe$_{3}$ & 192171 & 0.00 & 0.01 \\
        HgSnO$_{3}$ & 260029 & 0.00 & 0.03 \\
        TlInSe$_{2}$ & 180272 & 0.01 & 0.00 \\
        \bottomrule
        \end{tabular}
    \end{threeparttable}
    \label{tab:start_cmp_pbe}
\end{table}

A similar picture arises for the PBE calculations, as shown in Supplementary~Fig.~\ref{fig:start_cmp_pbe} and the associated Supplementary~Tab.~\ref{tab:start_cmp_pbe}.
Once again, discrepancies for Ag$_2$PdO$_2$, Cu$_2$GeSe$_3$, and HgSnO$_3$ correlate with the presence of additional semicore states ($4s4p$ for Ag, $3s3p$ for Cu, and $5s5p$ for Hg) in the NC pseudopotentials that are not included in the PAW pseudopotentials.
In the case of TlInSe$_2$, we are unsure what exactly causes the difference, as both the PAW and NC pseudopotentials contain the same valence and semicore orbitals.
We also performed an NC PBE calculation using an even finer k-point grid and still observed a very narrow band gap.
Once again, we conjecture that the missing norm-conservation condition in the PAW pseudopotentials, or differences in the plane-wave cutoff, may be responsible for the discrepancy.


\section*{Supplementary Note 2: Additional histograms}

\begin{figure}[ht]
    \centering
    \includegraphics{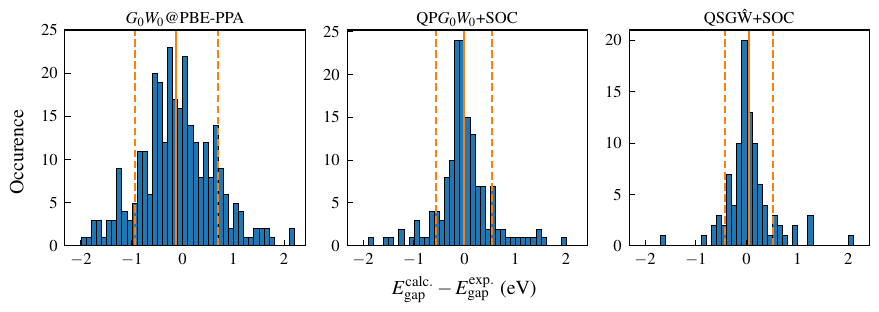}
    \caption{
    \textbf{Additional histograms showing the differences between calculated and experimental band gaps.}
    Each bin has a width of $0.1$~eV.
    The mean error (ME) is indicated by a solid orange line, while the standard deviation ($\sigma$) is indicated by dashed orange lines.
    }
    \label{fig:add_histo}
\end{figure}

In addition to Fig.~1 of the main text, Supplementary Fig.~\ref{fig:add_histo} shows further comparisons between theory and experiment as histograms.

\section*{Supplementary Note 3: MAPE discussion}

To further highlight why the MAPE is arguably an unsuitable metric for a band gap benchmark, we analyzed the distribution of the absolute percentage error (APE), defined as follows
\begin{equation*}
    \mathrm{APE}\left[E_\mathrm{gap}^\mathrm{exp.}; E_\mathrm{gap}^\mathrm{calc.}\right] = 100 \cdot \left| \frac{ E_\mathrm{gap}^\mathrm{exp.} - E_\mathrm{gap}^\mathrm{calc.}}{E_\mathrm{gap}^\mathrm{exp.}} \right|
\end{equation*}
between the experimental and QS$G\hat{W}$ band gaps.
The histogram of the APE, shown in Supplementary~Fig.~\ref{fig:ape_histo}, exhibits a pronounced right-skew with a long tail, similar to a log-normal distribution.
This behavior is inherent to the MAPE, since the values are bounded below by zero, while being unbounded above. 
This makes large over-predictions far more frequent than equally large under-predictions.
Consequently, a handful of extreme cases inflate the arithmetic mean to $29\%$, while the median remains modest at $7\%$, indicating that the "typical" error is much smaller than the mean suggests. 
Since the denominator is the experimental gap, $E_\mathrm{gap}^\mathrm{exp.}$, the composition of the dataset strongly influences the metric.
Materials with small $E_{\text{gap}}^{\text{exp}}$ contribute disproportionately, meaning that adding more narrow-gap semiconductors (or removing wide-gap insulators) can increase the overall MAPE, even if the absolute errors remain unchanged.
This fact is also emphasized by the observation in Supplementary~Table~\ref{tab:ape_outliers}, as the materials with the largest APE values all have small band gaps, and somewhat interestingly almost all contain antimony.

\begin{figure}[ht]
    \centering
    \includegraphics{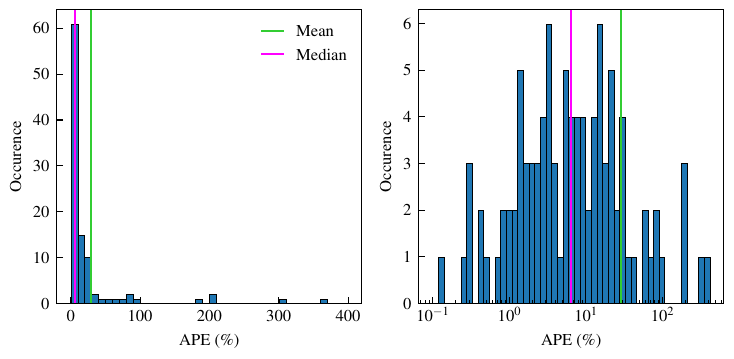}
    \caption{
    \textbf{Histograms showing the distribution of the absolute percentage error (APE) between the experimental and QS$\bm{G\hat{W}}$ band gaps.}
    The green line highlights the mean APE (MAPE), while the magenta line highlights the median APE.
    The right panel shows the same information as the left panel, but with a logarithmic abscissa.
    }
    \label{fig:ape_histo}
\end{figure}

\begin{table}[ht]
    \caption{
    \textbf{Materials with the highest APE between the experimental and QS$\bm{G\hat{W}}$ band gaps.}
    }
    \centering
    \begin{threeparttable}
        \begin{tabular}{lccccc}
        \toprule
        Material ~&~ ICSD-ID ~&~ QS$G\hat{W}$ (eV) ~&~ QS$G\hat{W}$+SOC (eV) ~&~ {Exp.} (eV) ~&~ MAPE (\%) \\
        \midrule
        YPtSb & 44970 & 0.83 & 0.58 & 0.16\tnote{a} & 419 \\
        PbTe & 63098 & 0.89 & 0.34 & 0.19\tnote{b} & 368 \\
        ScNiSb & 40296 & 0.44 & 0.40 & 0.11\tnote{a} & 300 \\
        InSb & 162196 & 0.74 & 0.51 & 0.24\tnote{b} & 208 \\
        YNiSb & 105331 & 0.54 & 0.50 & 0.18\tnote{a} & 200 \\
        \bottomrule
        \end{tabular}
        \begin{tablenotes}
            \item[a] High-temperature resistivity measurements, Ref.~\cite{Oestreich2003}.
            \item[b] Taken from Ref.~\cite{Madelung2004}.
        \end{tablenotes}
    \end{threeparttable}
    \label{tab:ape_outliers}
\end{table}


\section*{Supplementary Note 4: Additional calculations for \NoCaseChange{CuLaO$_2$}}

Since a QS$G\hat{W}$ calculation for CuLaO$_2$ using our benchmark workflow (see the Methods section in the main text) was beyond our computational capabilities, we approximate the QS$G\hat{W}$ band gap here using a reduced basis set and a truncated BSE transition space.
Here, we provide some details for completeness and reproducibility, but also as a suggested workaround for similar situations.

Starting from the benchmark basis set, we made the following adjustments:
(i) We removed "floating orbitals" from the basis set.
(ii) We reduced $\Sigma_\mathrm{cut}$ from $3$~Ry to $2$~Ry.
(iii) We reduced the plane-wave cutoff energies for the one- and two-particle basis sets in the $GW$ part of the \textsc{Questaal} code from $\texttt{gcutb}=4.0$ and $\texttt{gcutx}=3.3$ to $\texttt{gcutb}=2.8$ and $\texttt{gcutx}=2.3$, respectively.
We kept the DFT and $\Sigma$ k-point grids at the converged values from the benchmark calculations.
Since the BSE transition space suggested by our heuristic, as described in the Methods section in the main text, was quite large for CuLaO$_2$, i.e., $N_v = 11$ and
$N_c = 16$, we also use a truncated BSE transition space.
However, to still estimate a converged QS$G\hat{W}$ band gap, we iteratively increase the number of valence ($N_v$) and conduction ($N_c$) bands included in the BSE transition space. 
Starting from $N_v = N_c = 4$, we iteratively increased the number of valence and conduction bands included in the BSE transition space by two.
The BSE transition space used for the respective calculations is indicated by the following notation: QS$G\hat{W}@(N_v, N_c)$.
The results are listed in Supplementary~Tab.~\ref{tab:gap_conv}.
Although the band gap for QS$G\hat{W}@(8, 8)$ is not fully converged, it is a reasonable estimate.
Note, that the QS$G\hat{W}@(8, 8)$ took almost 20 days to complete using all $64$ cores of an AMD~Milan~7763, with a peak memory usage of around $700$~GB.
These calculations indicate that the BSE transition space produced by our heuristic will likely be fully converged.

\begin{table}[ht]
    \caption{
    \textbf{Band gap of CuLaO$_2$ calculated using different methods.}
    All band gaps are given in eV and were estimated from band structure calculations, cf.~Supplementary~Note~6.
    The values in the "Benchmark Setting" row come directly from the benchmark dataset. The values in the "Reduced Setting" row are calculated using a reduced basis set and a reduced BSE transition space, as described in the text above.
    The notation is explained in the text.
    }
    \centering
    \begin{threeparttable}
        \begin{tabular}{lcccccc}
        \toprule
        & LDA & QS$GW$ & QS$G\hat{W}@(4, 4)$ & QS$G\hat{W}@(6, 6)$ & QS$G\hat{W}@(8, 8)$ & Exp.\\ \midrule
        Benchmark Setting & 2.48 & 5.09 & --- & --- & --- & \multirow{2}{*}{2.34\tnote{a}} \\
        Reduced Setting & 2.48 & 4.93 & 4.53 & 4.46 & 4.36 & \\
        \bottomrule
        \end{tabular}
        \begin{tablenotes}
            \item[a] Diffuse reflectance spectra, Ref.~\cite{Koriche2006}.
        \end{tablenotes}
    \end{threeparttable}
    \label{tab:gap_conv}
\end{table}

\begin{figure}[ht]
    \centering
    \includegraphics{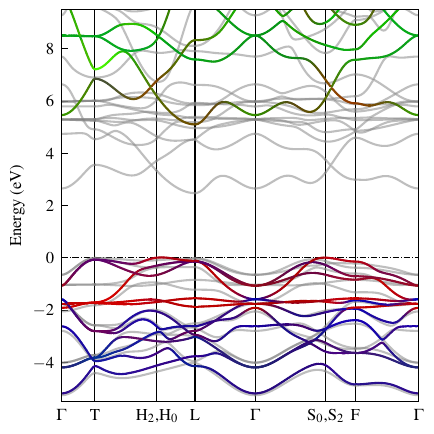}
    \caption{
    \textbf{Element-resolved QS$\bm{G}\bm{W}$ band structure of CuLaO$_2$ calculated using the benchmark settings.}
    The symmetry path was taken from Ref.~\cite{Hinuma2017}. 
    The contribution of Cu, La, and O states to any given band is color-coded as red, green, and blue components of its hue.
    Gray bands show the LDA band structure.
    }
    \label{fig:bs_culao2}
\end{figure}

As already described in the main text, the element-resolved QS$GW$ band structure of CuLaO$_2$ shown in Supplementary~Fig.~\ref{fig:bs_culao2} reveals an almost dispersionless valence band comprised of Cu states, with minor contributions from O states.
The lowest conduction band appears to be a hybridization of Cu and La states, with some minor contributions from O around the T point.


\section*{Supplementary Note 5: Sanitized benchmark dataset metrics}

Analogous to the main text, we report all error metrics for all selected computational methods evaluated on the sanitized version of the benchmark dataset of Borlido \textit{et al.}~\cite{Borlido2019, Borlido2020}.
To obtain the sanitized version of the benchmark dataset, we removed the following compounds from the original dataset: CuLaO$_2$, Cu$_2$S, Cu$_2$Se, MgCl$_2$, PbF$_2$, and SrF$_2$.
Additionally, we updated the experimental references for BeSe, BaO, and ScN, adopting band gap values of $3.8$~eV measured by Eich \textit{et al.}~\cite{Eich2006}, $4.1$~eV measured by Saum \textit{et al.}~\cite{Saum1959}, and $1.1\pm0.1$~eV measured by Grümbel \textit{et al.}~\cite{Gruemble2024}, respectively.
The reasons for these changes are explained in detail in the main text.
The metrics for mBJ and HSE06 have remained essentially unchanged or improved minimally, as have those for $G_0W_0$@LDA-PPA and $G_0W_0$@PBE-PPA.
Substantial improvements in most metrics are observed for the QP$G_{0}W_{0}$, QS$GW$, and QS$G\hat{W}$ calculations.

\begin{table*}[ht]
    \caption{
    \textbf{Error metrics for all investigated calculation methods, evaluated on the sanitized version of the benchmark dataset.}
    As in Tab.~I of the main text, the following metrics are given for the calculated band gaps using the methods described in the text: Dataset size ($n$), mean error (ME), standard deviation of the error ($\sigma$), mean absolute error (MAE), root mean square error (RMSE), and mean absolute percentage error (MAPE).
    The error is defined as $E_{\mathrm{gap}}^{\mathrm{calc.}} - E_{\mathrm{gap}}^{\mathrm{exp.}}$.
    The metrics in the upper block are based on all available calculations for each method, as indicated by the $n$ row. 
    The lower block, on the other hand, restricts the comparison to the 90 materials for which data is available for all methods.
    Note that the MAPE has a tendency to preference methods which systematically underestimate band gaps, see explanation in the main text and Supplementary~Note~3.
    The best method for each metric is highlighted in bold.
    \\~
    }
    \centering
    \squeezetable
    \scriptsize	
    \begin{tabular}{lccccccccc}
    \toprule
     & mBJ & HSE06 & $G_{0}W_{0}$@LDA-PPA & $G_{0}W_{0}$@PBE-PPA & QP$G_{0}W_{0}$ & QP$G_{0}W_{0}$+SOC & QS$GW$ & QS$G\hat{W}$ & QS$G\hat{W}$+SOC \\
    \midrule
    $n$ & 463 & 464 & 279 & 294 & 148 & 145 & 146 & 97 & 92 \\
    ME (eV) & -0.21 & -0.09 & -0.02 & -0.11 & 0.09 & \textbf{0.00} & 0.56 & 0.12 & 0.03 \\
    $\sigma$ (eV) & 0.67 & 0.84 & 0.74 & 0.82 & 0.43 & 0.44 & 0.48 & \textbf{0.32} & 0.34 \\
    MAE (eV) & 0.48 & 0.51 & 0.52 & 0.59 & 0.31 & 0.30 & 0.57 & \textbf{0.24} & \textbf{0.24} \\
    RMSE (eV) & 0.70 & 0.84 & 0.74 & 0.82 & 0.44 & 0.44 & 0.73 & 0.34 & \textbf{0.33} \\
    MAPE (\%) & 29 & 30 & 39 & 40 & 31 & 25 & 48 & 29 & \textbf{20} \\
    \midrule
    \\[-3mm]
    $n$ & \multicolumn{9}{c}{Materials for which experimental and computational data are available for all methods: 90} \\
    ME (eV) & -0.42 & -0.67 & -0.15 & -0.24 & 0.04 & -0.05 & 0.53 & 0.11 & \textbf{0.01} \\
    $\sigma$ (eV) & 0.65 & 1.17 & 0.54 & 0.66 & 0.38 & 0.37 & 0.41 & \textbf{0.32} & \textbf{0.32} \\
    MAE (eV) & 0.56 & 0.82 & 0.44 & 0.55 & 0.26 & 0.26 & 0.54 & 0.24 & \textbf{0.23} \\
    RMSE (eV) & 0.77 & 1.35 & 0.56 & 0.70 & 0.38 & 0.37 & 0.67 & 0.34 & \textbf{0.32} \\
    MAPE (\%) & 36 & 39 & 36 & 38 & 29 & 20 & 43 & 30 & \textbf{20} \\
    \bottomrule
    \end{tabular}
    \label{tab:clean_metrics}
\end{table*}

\section*{Supplementary Note 6: Band gap estimation methods}

This short note addresses the issue of band gap estimations in the context of automated workflows and high-throughput calculations. 
Band gaps can be obtained either using a regular three-dimensional k-point grid ($E_\mathrm{gap}^{\mathbf{k}_\mathrm{DFT}}$), or from separate band structure calculations on high-symmetry lines that connect special points in the Brillouin Zone ($E_\mathrm{gap}^{\mathbf{k}_\mathrm{sym}}$).
Tacitly, we assume in the following comparison that approximately the same reasonable effort is spent on either approach. 
Obviously, $E_\mathrm{gap}^{\mathbf{k}_\mathrm{DFT}} \leq E_\mathrm{gap}^{\mathbf{k}_\mathrm{sym}}$ in the limit of infinitely many k-points as $\mathbf{k}_\mathrm{DFT}$ will densely cover all symmetry axes, whereas $\mathbf{k}_\mathrm{sym}$ will never come close to whole regions of "symmetry-free" points in reciprocal space. 
There is no theorem stating that band extrema must occur on special points or symmetry axes connecting them.   
As mentioned in the main text, we extract band gaps from the all-electron LDA DFT, QP$G_{0}W_{0}$, QS$GW$, and QS$G\hat{W}$ calculations performed using the \textsc{Questaal} code by combining both Brillouin zone sampling strategies:
\begin{equation*}
    E_\mathrm{gap} = \min\left(E_\mathrm{gap}^{\mathbf{k}_\mathrm{DFT}}, E_\mathrm{gap}^{\mathbf{k}_\mathrm{sym}} \right)
\end{equation*}

\begin{figure}[ht]
    \centering
    \includegraphics{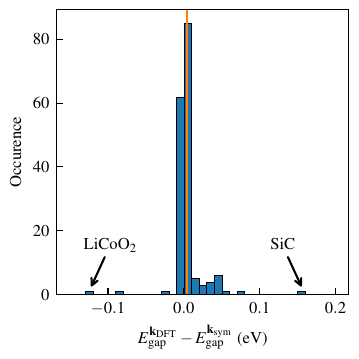}
    \caption{
    \textbf{Histogram showing the difference between the \textsc{Questaal} LDA band gaps obtained using two different Brillouin zone sampling strategies.}
    We compare band gaps calculated using a regular three-dimensional k-point grid converged with respect to the total energy per atom, i.e., ${E_\mathrm{gap}^{\mathbf{k}_\mathrm{DFT}}}$, with those from band structure calculations along high-symmetry lines, i.e., ${E_\mathrm{gap}^{\mathbf{k}_\mathrm{sym}}}$.
    Each bin has a width of $0.01$~eV.
    The mean error (ME) is indicated by a solid orange line.
    Arrows mark materials for which the difference is large.
    }
    \label{fig:gap_estim}
\end{figure}

Supplementary~Figure~\ref{fig:gap_estim} shows the difference between LDA band gaps obtained from a regular three-dimensional k-point grid and those obtained from band structures.
The average (signed) difference is only $4$~meV (orange line), and the mean absolute difference is $7$~meV, indicating that the two strategies are consistent to within a few meV for the vast majority of materials calculated using our workflow.
However, we note that there are outliers. 

Materials where $E_\mathrm{gap}^{\mathbf{k}_\mathrm{DFT}} - E_\mathrm{gap}^{\mathbf{k}_\mathrm{sym}} > 0$ are attributed to the presence of an often indirect band gap along symmetry lines between high-symmetry points that are not adequately covered by our regular three-dimensional k-point grids.
As shown in Supplementary~Fig.~\ref{fig:gap_estim}, materials with $E_\mathrm{gap}^{\mathbf{k}_\mathrm{DFT}} - E_\mathrm{gap}^{\mathbf{k}_\mathrm{sym}} < 0$ are significantly less common in the benchmark dataset.
In these cases, the band extrema are located at "symmetry-free" k-points in the Brillouin zone that are "far" from any symmetry lines.
Note that in our $G_{0}W_{0}$-PPA workflow (see the Methods section in the main text), we converged the three-dimensional DFT k-point grid with respect to the band gap and thus no band structures are required there.

\begin{figure}[ht]
    \centering
    \includegraphics{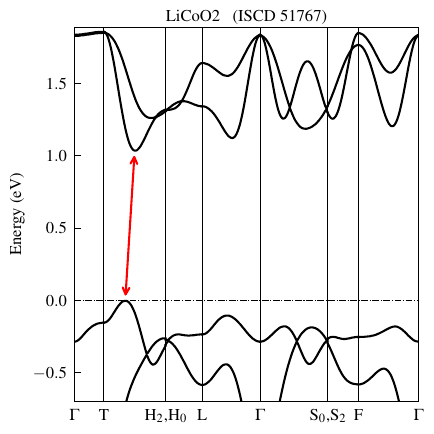}
    \caption{    
    \textbf{LDA band structure of R$\bm{\overline{3}}$m LiCoO$_2$.}
    The symmetry path was taken from Ref.~\cite{Hinuma2017}. 
    A red arrow highlights the indirect band gap.
    }
    \label{fig:bs_licoo2}
\end{figure}

We decided to take a closer look at the extreme outlier LiCoO$_2$ (space group R$\overline{3}$m) because we found it intriguing that its band gap is strongly overestimated by a band structure calculation.
This has also been noted before in Ref.~\cite{Radha2021}.
Examining the LDA band structure of LiCoO$_2$ shown in Supplementary~Fig.~\ref{fig:bs_licoo2}, which was calculated using a symmetry path taken from Ref.~\cite{Hinuma2017}, we find an indirect band gap of $1.039$~eV along the T-H$_2$ symmetry line.
However, we find a smaller direct band gap of $0.918$~eV using a regular three-dimensional DFT k-point grid that was converged with respect to the total energy per atom ($8 \times 8 \times 8$ k-point grid).
Using a dense $24 \times 24 \times 24$ ($48 \times 48 \times 48$) k-point grid, we find an even smaller indirect band gap of $0.892$~eV ($0.880$~eV).
The exact k-points of these band gaps are given in Supplementary~Tab.~\ref{tab:licoo2_gap}.

\begin{table}[ht]
    \caption{
    \textbf{LDA band gap of R$\bm{\overline{3}}$m LiCoO$_2$ calculated with increasingly fine k-point grids.}
    The exact k-points for the valence band maximum (VBM) and conduction band minimum (CBM) are given in reciprocal coordinates.
    The k-plane (i) is defined in the text below, cf.~Supplementary~Fig.\ref{fig:vbm_cbm_map}.
    }
    \centering
    \begin{tabular}{@{}cccccc@{}}
    \toprule
    & Band structure & $8 \times 8 \times 8$ & $24 \times 24 \times 24$ & $48 \times 48 \times 48$ & k-plane (i) \\ \midrule
    $E_\mathrm{gap}$ (eV)     & 1.039 & 0.918 & 0.892  & 0.880 & 0.880   \\
    $\mathbf{k}_\mathrm{VBM}$ & $(0.608, 0.392, 0.500)$ & $(0.500, 0.250, 0.250)$ & $(0.667, 0.458, 0.667)$ & $(0.667, 0.458, 0.667)$ & $(0.667, 0.458, 0.667)$ \\
    $\mathbf{k}_\mathrm{CBM}$ & $(0.657, 0.343, 0.500)$ & $(0.500, 0.250, 0.250)$ & $(0.250, 0.250, 0.542)$ & $(0.271, 0.271, 0.542)$ & $(0.271, 0.271, 0.541)$ \\ \bottomrule
    \end{tabular}
    \label{tab:licoo2_gap}
\end{table}

To investigate the topology of the valence band maximum (VBM) and the conduction band minimum (CBM), we calculated their dispersions using very dense $101 \times 101$ two-dimensional k-point grids in selected planes of the Brillouin zone.
(i) Data‑driven plane: spanned by $\mathbf{k}_\mathrm{T}\rightarrow\mathbf{k}_\mathrm{H_2}$ and $\mathbf{k}_\mathrm{T}\rightarrow\mathbf{k}_\mathrm{VBM/CBM}$, where $\mathbf{k}_\mathrm{VBM/CBM}$ were identified from an LDA calculation on a $48 \times 48 \times 48$ k-point grid.
(ii) Symmetry‑guided plane: spanned by $\mathbf{k}_\mathrm{T}\rightarrow\mathbf{k}_\mathrm{H_2}$ and $\mathbf{k}_\mathrm{T}\rightarrow\mathbf{k}_\mathrm{S_4/F}$, i.e., vectors to the nearest high‑symmetry points adjacent to the data‑driven planes.
The VBM and CBM along the chosen k-point planes are visualized in Supplementary~Fig.~\ref{fig:vbm_cbm_map}.
Interestingly, we find that the indirect band gap of $0.889$~eV obtained from planes (ii) almost matches the band gap of $0.880$~eV from planes (i).
The positions of the VBM and CBM within the Brillouin zone are given in Supplementary~Tab.~\ref{tab:licoo2_gap}.
At this point, we note that, as Radha \textit{et al.}~\cite{Radha2021} noticed, the exact positions of the VBM and CBM depend on the level of theory of the calculation, cf.~Tab.~1 in Ref~\cite{Radha2021}.
This example clearly shows how difficult it is to obtain accurate band gaps, a problem that plagues high-throughput workflows like ours.
Nevertheless, even in this extreme case, the difference in the band gap between a "normal" k-point grid and an extremely dense one is only $40$~meV.

\begin{figure}[ht]
    \centering
    \includegraphics{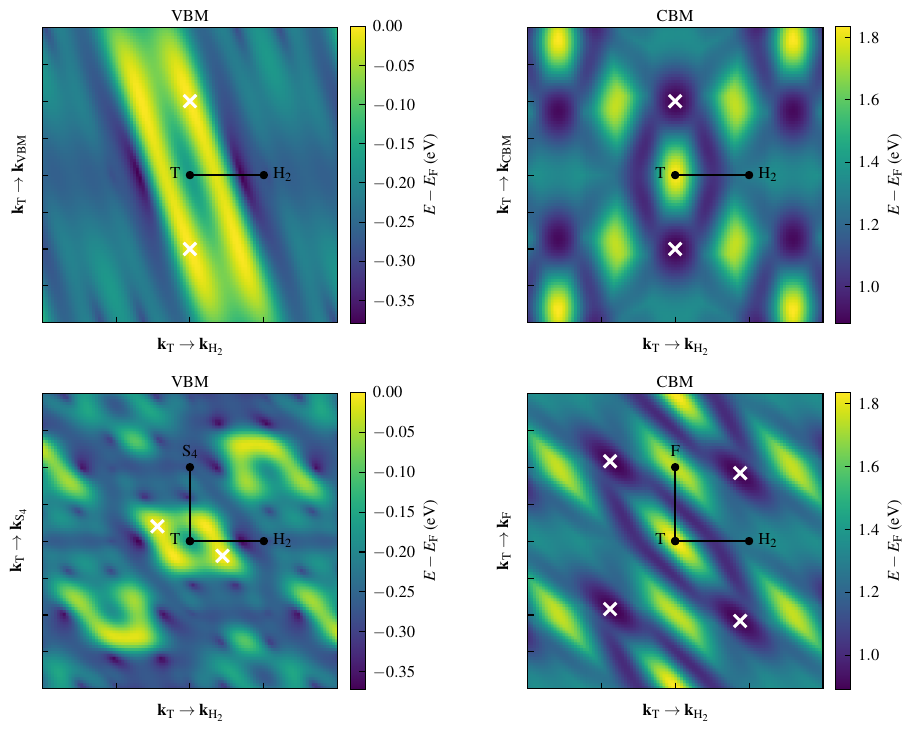}
    \caption{    
    \textbf{Highest valence band and lowest conduction band of R$\bm{\overline{3}}$m LiCoO$_2$ calculated in the LDA along different k-point planes in the Brillouin Zone.}
    The text above explains how the planes were selected.
    Note that the abscissa and ordinate are not based on orthogonal vectors.
    Symmetry lines are highlighted in black, and extrema are marked with a white cross.
    Figure~18 in Ref.~\cite{Hinuma2017} shows an illustration of a hexagonal rhombohedral ($h$sR) Brillouin zone including all high-symmetry points.
    }
    \label{fig:vbm_cbm_map}
\end{figure}

In conclusion, we would like to emphasize that these findings, together with Supplementary~Note~1, have implications that extend far beyond the scope of this benchmark.
Most large materials science databases containing band gaps and other properties derived from electronic structures are calculated using the \textsc{VASP} code \cite{Kresse1996, Kresse1999} utilizing PAW pseudopotentials \cite{Bloechl1994}.
The findings of Supplementary~Note~1 tentatively suggest that some materials in these databases may be misclassified, with metals labeled as narrow-gap semiconductors and, conversely, narrow-gap semiconductors labeled as metals.
This issue is likely exacerbated by the widespread use of fixed k-point grid densities in the calculations used to create most large databases.
We suspect that these issues are currently hindering the performance of machine learning models that have been trained to predict the band gap using these databases.
Future research should address this issue.


\bibliography{literature}